\documentclass[10pt]{article}
\usepackage[affil-it]{authblk}
\usepackage{fullpage}
\usepackage{url}
\usepackage{amsmath}
\usepackage{amsfonts}
\usepackage{graphicx}
\usepackage{subcaption}
\usepackage{algorithm}
\usepackage{algorithmic}
\usepackage{hyperref}
\usepackage[UKenglish]{babel}
\DeclareMathOperator*{\argmin}{\arg\!\min}

\usepackage{natbib}

\begin{document}
\title{Quantitative cone-beam CT reconstruction with \\ polyenergetic scatter model fusion}

\author{Jonathan H. Mason\textsuperscript{1}%
	\thanks{Electronic address: \texttt{j.mason@ed.ac.uk}; Corresponding author} }
\author{Alessandro Perelli\textsuperscript{1}}\author{William H. Nailon\textsuperscript{1,2}}\author{Mike E. Davies\textsuperscript{1}}
\affil{\textsuperscript{1}School of Engineering, Institute for Digital Communications, \\University of Edinburgh, Edinburgh, EH9 3JL, UK}
\affil{\textsuperscript{2}Department of Oncology Physics, Edinburgh Cancer Centre, \\Western General Hospital, Edinburgh, EH4 2XU, UK}

\maketitle
\begin{abstract}
Scatter can account for large errors in cone-beam CT (CBCT) due to its wide field of view, and its complicated nature makes its compensation difficult. Iterative polyenergetic reconstruction algorithms offer the potential to provide quantitative imaging in CT, but they are usually incompatible with scatter contaminated measurements. In this work, we introduce a polyenergetic convolutional scatter model that is directly fused into the reconstruction process, and exploits information readily available at each iteration for a fraction of additional computational cost. We evaluate this method with numerical and real CBCT measurements, and show significantly enhanced electron density estimation and artifact mitigation over pre-calculated fast adaptive scatter kernel superposition (fASKS). We demonstrate our approach has two levels of benefit: reducing the bias introduced by estimating scatter prior to reconstruction; and adapting to the spectral and spatial properties of the specimen. 
\end{abstract}

\section{Introduction}
X-ray scatter is a large source of errors in CT, where scattered X-rays corrupt the line of sight attenuation models used during reconstruction. In CBCT, due to its large field of view, the magnitude of these interfering X-rays is commonly of the same order of magnitude as the signal of interest and can even be considerably higher \citep{Siewerdsen2001}. With this, artifacts and inaccuracies in the reconstruction are inevitable unless it can be correctly compensated. In \citet{Mason2017b}, we introduced a polyenergetic quantitative reconstruction technique allowing the direct inference of electron density --- known as Polyquant. This method can incorporate a prior estimate of the additive scatter, but this does not exploit any of the polyenergetic information or indeed the electron density available during iterations.

The problem then remains of how one should generate this estimate of scatter, which is an active area of research with many possible approaches \citep{Ruhrnschopf2011b}. A fundamental reason why estimating scatter is difficult is that unlike modelling attenuation, requiring a single path from source to detector for each measurement, one must consider every possible path a photon can take through various numbers of scattering events to get the full picture. This will therefore not only be dependent on the attenuating materials and intensity along a single pencil beam, but it will be dependent on the full projection fluence and the complete structure of the specimen. Not only would calculating all these paths be computationally exhaustive, but one does not even have knowledge of the structure of the specimen prior to reconstruction, since this is the task of the reconstruction itself. This therefore presents a dependency problem, where one requires the image to estimate scatter, but cannot form the image accurately because of scatter contamination.

To overcome this limitation, a compelling approach is to integrate the scatter estimation and reconstruction processes, so that they are performed simultaneously. This notion is exploited to some degree by efficient Monte Carlo scatter estimation methods such as \citet{Poludniowski2009a,Zbijewski2006,Sisniega2015,Xu2015}, where one requires a preliminary reconstruction to estimate the scatter, which can then be iteratively refined. Another approach in \citet{Lee2015} used an analytic single scatter model to generate scatter estimates during reconstruction. A drawback of either Monte Carlo or analytic scatter models is their computational cost, which would make the already high cost of iterative reconstruction huge.

There are also hardware approaches to scatter estimation, which usually allow very cheap computational estimation, but typically at the loss of detector utilisation, unless additional acquisitions are taken \citep{Ruhrnschopf2011b}. One such approach proposed by \citet{Siewerdsen2006} directly measures the scatter that hits the edge of the detector, in the collimator shadow, and interpolates this over an entire projection, through low-order polynomial fitting and filtering. With this, one has a trade off between the field of view size, and accuracy at the center of the detector. A similar method, beam stop array (BSA), measures the scatter behind an array of beam blockers distributed over the entire detector area \citep{Ning2004}, and uses interpolation between these sparse points to form the estimate. By moving the beam blockers throughout the acquisition, then separate scans need not be taken \cite{Zhu2005}. In this work, we will be primarily be focussing on methods that utilize the entire detector area and with no additional hardware, but we will compare against a beam stop array approach in Section~\ref{sec:bsa}.

Another family of scatter estimation approaches that are purely measurement based are convolutional methods, known as scatter kernel superposition (SKS) \citep{Sun2010} or beam scatter kernels (BSK) \citep{Ruhrnschopf2011b}. Here, the scatter is estimated as a convolution between a stationary point spread function, usually calculated or measured for a slab of material, and a forward scatter factor \citep{Ohnesorge1999} derived from the small angle scatter magnitude hitting the line-of-sight detector element. SKS methods are typically very fast, due to their ability to exploit the fast Fourier transform (FFT) based calculation of convolution, but are naturally limited in their inability to account for inhomogeneities in the scatter paths. These methods may however be supplemented with heuristic perturbations to compensate for the most significant inhomogeneous effects such as fringing \citep{Maltz2008}, varying thickness \citep{Sun2010}, effects at the edge of an object \citep{Sun2010}, scatter from the couch \citep{Sun2011}, and more general spatial inhomogeneities \citep{Sun2014}.

From a preliminary study comparing scatter estimation methods when used with statistical iterative reconstruction \citep{Mason2017d}, we found that SKS approaches matched the performance of sub-sampled Monte Carlo estimates, whilst being significantly faster. For this reason, we focus on the integration of SKS models into the polyenergetic reconstruction algorithm, Polyquant, and see if their performance can be enhanced further from information available during the iterations.

In this work, we derive a polyenergetic SKS (PolySKS) model in terms of attenuation and electron density projections, include perturbations to account for the specimen position and inhomogeneities in scatter paths, and integrate this into the Polyquant framework. We denote this integrated reconstruction method as Polyquant--PolySKS. While polyenergetic kernel convolution has been used for radiotherapy dose calculation \citep{Papanikolaou1993}, we provide the analysis and derivation in this article for CT imaging. We then propose a fast algorithm for its implementation using accelerated ordered subsets as in \citet{Kim2015a}, able to include sparsity exploiting regularization such as total variation (TV). Finally, we demonstrate the method on both numerical and real CBCT data, and analyse its performance in terms of relative electron density accuracy.

\section{Background}
\subsection{X-ray Attenuation Model and Polyquant}
In this section, we will review X-ray attenuation and the Polyquant model as presented in \citet{Mason2017b} for direct electron density imaging.

Considering a polyenergetic X-ray source, the intensity after transmission through a specimen can be modelled using an approximate Poisson distribution in the discrete domain \citep{Chang2014} as
\begin{equation} \label{equ:poly-poiss}
y_i \sim \operatorname{Poisson}\left\{\sum_{j=1}^{N_\xi}b_i(\xi_j)\exp\left(-[\boldsymbol{\Phi}\boldsymbol{\mu}(\xi_j)]_i\right)+s_i \right\} \mbox{ for }  i=1,...,N_\mathrm{ray},
\end{equation}
where $N_\xi$ are the number of energies, $\boldsymbol{b}\in\mathbb{R}^{N_\mathrm{ray}}$ is the spectrally dependent source intensity, $\boldsymbol{\Phi}\in\mathbb{R}^{N_\mathrm{ray}\times N_\mathrm{vox}}$ is a system matrix describing the line-of-sight path of each ray, $\boldsymbol{\mu}(\xi)\in\mathbb{R}^{N_\mathrm{vox}}$ is the energy dependent X-ray attenuation, and $\boldsymbol{s}\in\mathbb{R}^{N_\mathrm{ray}}$ is the expectation of scatter reaching the detector. $N_\mathrm{ray}$ and $N_\mathrm{vox}$ represent the number of measurements (rays) and voxels in the discretized volume respectively.

The Polyquant model allows the parameterization of this attenuation process as a piecewise linear function of the electron density. This is shown in \citet{Mason2017b} to fit well to a range of biological tissues such as soft tissues and bone, but also metallic materials for implants such as titanium. This linear fit is expressed as
\begin{equation} \label{equ:gen_fit}
\hat{\boldsymbol{\mu}}(\boldsymbol{\rho}_e,\xi) = \sum_{l=1}^{N_f}\boldsymbol{f}_l\odot(\alpha_l(\xi)\boldsymbol{\rho}_e+\beta_l(\xi)), 
\end{equation}
where $\odot$ represents an element-wise multiplication, $N_f$ is the number of linear intervals --- typically $N_f=2$ for biological tissues and $N_f=3$ with the inclusion of metal --- and $\boldsymbol{\rho}_e\in\mathbb{R}^{N_\mathrm{vox}}$ is the relative electron density defined as \citep{Schneider1996}
\begin{equation} \label{equ:red}
\rho_e = \frac{\rho N_g}{\rho_\mathrm{water}N_{g,\mathrm{water}}},
\end{equation}
where $\rho$ is the mass density, $N_g$ is the electrons per unit volume, and the subscript indicates associated values for liquid water at standard temperature and pressure. $\alpha$ and $\beta$ are fitted parameters to form a connected piecewise linear function to X-ray attenuation data such as from \citet{icrp2002}. To ensure vacuum results in no attenuation, the fit is constrained such that $\beta_1=0$. The thresholding vector $\boldsymbol{f}_l$ is a mask to isolate voxels in the image corresponding to the $l^\mathrm{th}$ section of the linear fit as
\begin{equation} \label{equ:threshold}
f_{l,i} = 
\begin{cases} 
1 & \text{if } k_{l-1} \leq \rho_{e,i} < k_l\\
0       & \text{otherwise}
\end{cases}
\mbox{ for }  l=1,\dots,N_f; \mbox{ for }  i=1,\dots,N_\mathrm{ray},
\end{equation}
where $k$ are the `knee' transition points between linear intervals, with $k_0 = 0$, and (\ref{equ:threshold}) is updated at each iteration. By using the substitution
\begin{equation} \label{equ:psi}
\psi_i(\boldsymbol{\rho}_e,\xi) \equiv b_i(\xi)\exp\left(-[\boldsymbol{\Phi}\boldsymbol{\hat{\mu}}(\boldsymbol{\rho}_e,\xi)]_i\right) \mbox{ for }  i=1,...,N_\mathrm{ray},
\end{equation}
the negative log-likelihood (NLL) of the noise model in (\ref{equ:poly-poiss}) can then be expressed in terms of the relative electron density as
\begin{equation} \label{equ:nll}
\mathrm{NLL}(\boldsymbol{\rho}_e;\boldsymbol{y}) =
\sum_{i=1}^{N_\mathrm{ray}}\sum_{j=1}^{N_\xi}\psi_i(\boldsymbol{\rho}_e,\xi_j)+s_i - y_i\log\left(\sum_{j=1}^{N_\xi}\psi_i(\boldsymbol{\rho}_e,\xi_j)+s_i\right),
\end{equation}
which is differentiable, and can be minimized through gradient descent with appropriate regularization in an algorithm such as the `Prox-Polyquant' given in \citet{Mason2017b}.

One significant discrepancy between the model in (\ref{equ:poly-poiss}) and reality however, is that the scatter term $\boldsymbol{s}$ is not independent. Instead it is highly dependent on the attenuation of the specimen $\boldsymbol{\mu}$, the intensity and spectrum of the source $\boldsymbol{b}$, the system geometry $\boldsymbol{\Phi}$, and also the electron density $\boldsymbol{\rho}_e$ as this determines the probability of a scatter event to occur \citep{Jackson1981}. With this, a more statistically meaningful expression for the transmission model is
\begin{equation} \label{equ:scat-poiss}
y_i \sim \operatorname{Poisson}\left\{\sum_{j=1}^{N_\xi}b_i(\xi_j)\exp\left(-[\boldsymbol{\Phi}\boldsymbol{\mu}(\xi_j)]_i\right)+s_i(\boldsymbol{b},\boldsymbol{\Phi},\boldsymbol{\mu},\boldsymbol{\rho}_e) \right\} \mbox{ for }  i=1,...,N_\mathrm{ray}.
\end{equation}
Due to this dependency, an appropriate model of scatter is required to perform Polyquant like reconstruction on the associated NLL function. This is what we develop in Section~\ref{sec4:poly_scat_model}.

\subsection{Convolutional Scatter Estimation} \label{sec:conv_classical}
\begin{figure}[!htb]
	\centering
	\includegraphics[width=0.9 \textwidth]{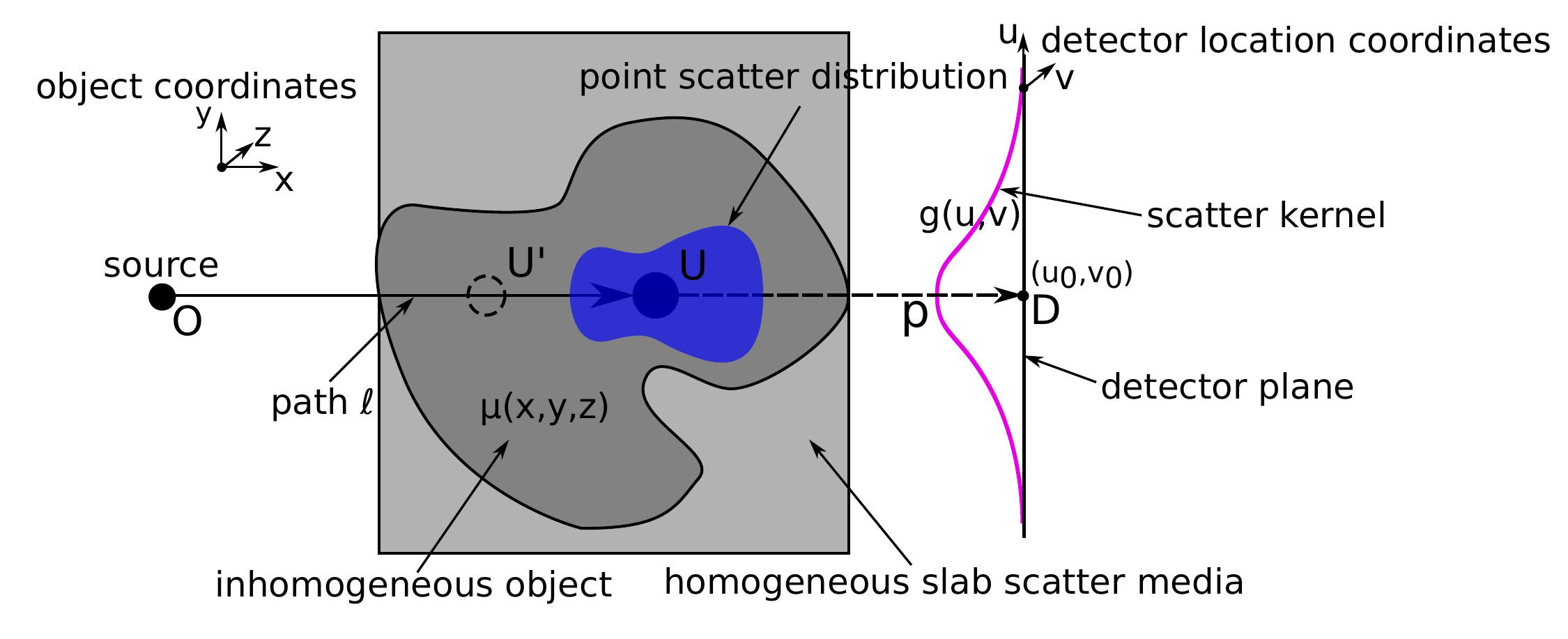}
	\caption{Scatter diagram showing homogeneous slab media and coordinate systems of object and detector along with point scatter distribution for forward scatter derivation.}
	\label{fig:scatter_deriv}
\end{figure}

A simple form of an SKS convolutional scatter model is represented in the discrete domain as \citep{Ohnesorge1999}
\begin{equation} \label{equ:simple_conv}
\boldsymbol{s}\{k\}(u,v) = \boldsymbol{p}\{k\}(u,v)*\boldsymbol{g}(u,v) \mbox{ for }  k=1,...,N_\mathrm{proj},
\end{equation}
where we use the notation $\{k\}(u,v)$ to indicate the $k$\textsuperscript{th} projection, with coordinates in the 2D detector plane $(u,v)$ as in Figure~\ref{fig:scatter_deriv}, $N_\mathrm{proj}$ is the number of projections, and $*$ represents a 2D convolution across the $(u,v)$ dimensions. With this, the scatter is modelled as some intensity function $\boldsymbol{p}$ filtered by a spatially invariant kernel $\boldsymbol{g}$. The assumption implied by this spatial invariance is that the scatter paths are through some homogeneous slab media, instead of the true inhomogeneous object, as is illustrated in Figure~\ref{fig:scatter_deriv}.

The scatter kernel function $\boldsymbol{g}$ in (\ref{equ:simple_conv}) often takes the form of the sum of two 2D Gaussian functions \citep{Ohnesorge1999,Sun2010} as 
\begin{equation}
g(u,v) = a_\mathcal{N}\exp\left(-\frac{(u-u_0)^2+(v-v_0)^2}{c_\mathcal{N}^2}\right) + a_\mathcal{B}\exp\left(-\frac{(u-u_0)^2+(v-v_0)^2}{c_\mathcal{B}^2}\right),
\end{equation}
where $(u_0,v_0)$ are the coordinates at the center of the detection plane, and $a$ and $c$ are the amplitude and width of the two Gaussian functions. We use the subscripts $\mathcal{N}$ and $\mathcal{B}$ to indicate the narrow and broad functions respectively.

The scatter intensity function $\boldsymbol{p}$ in (\ref{equ:simple_conv}) has some physical meaning, as it is the amount of scattered photons arriving along the initial trajectory of the ray. With reference to Figure~\ref{fig:scatter_deriv}, the derivation of this factor according to \citet{Ohnesorge1999} may be made in the continuous domain for a monoenergetic beam and ignoring multiple scattering. To begin with, the photon flux at point $U$ after travelling through the object is expressed as
\begin{equation}
I_U = I_O\exp\left(-\int_O^U \boldsymbol{\mu}(\ell) \, d\ell \right).
\end{equation}
At this point, photons will have a probability of scattering at any angle by electrons in the object. We will denote the proportion of photons scattered at an angle of $0^\circ$ as $K\mu(U)$, which is assumed proportional to the attenuation coefficient at that point. In reality, there will be a dependence on the spectrum of the source and the electron density, and we will explore this when developing the PolySKS model in Section~\ref{sec4:poly_scat_model}. For this to be detected at the point $D$, with coordinates $(u_0,v_0)$ in this case, it must travel through the rest of the object. The expectation of detected scatter from this single point is therefore
\begin{equation} \label{equ:scat_prob}
I_D = I_UK\mu(U)\exp\left(-\int_U^D \boldsymbol{\mu}(\ell) \, d\ell \right).
\end{equation}
The total scatter factor is the summation of all the individual scatter factors from $O$ to $D$, which can be expressed as
\begin{align} \label{equ:for_deriv_mono}
p &= \int_O^D I_O\exp\left(-\int_O^{U'} \boldsymbol{\mu}(\ell) \, d\ell \right)K\mu(U')\exp\left(-\int_{U'}^D \boldsymbol{\mu}(\ell) \, d\ell \right) \,dU' \nonumber \\
&= I_O\exp\left(-\int_O^D \boldsymbol{\mu}(\ell) \, d\ell \right) K \int_O^D\mu(\ell) \, d\ell.
\end{align}

By now adopting discrete notation, this may be expressed with respect to the system matrix $\boldsymbol{\Phi}$ as
\begin{equation} \label{equ:classic_for}
p_i = KI_\mathrm{in}\exp(-[\boldsymbol{\Phi\mu}]_i)[\boldsymbol{\Phi\mu}]_i \mbox{ for }  i=1,...,N_\mathrm{ray},
\end{equation}
where $I_\mathrm{in} = \sum_{j=1}^{N_\xi}b(\xi)$ is the total incident flux of the source. Now with (\ref{equ:classic_for}), it is possible to calculate the scatter factor at each iteration according to this monoenergetic model, and we will explore this concept with the Int-fASKS method tested in the experimental section. In practice however, one often performs the estimation as a preprocessing operation before reconstruction, when one has no knowledge of $\boldsymbol{\mu}$. By assuming the linearity from a monoenergetic source, an estimate is possible through calculating
\begin{equation} \label{equ:pre_forward}
p_i = Ky_i^{h_1}\left[-\log\left(\frac{y_i}{I_\mathrm{in}}\right)\right]^{h_2} \mbox{ for }  i=1,...,N_\mathrm{ray},
\end{equation}
where the empirical parameters $h_1$ and $h_2$ provide a better fit to a real source, and selected to match observation \citep{Ohnesorge1999}. They may be fitted by using a non-linear least squares approach such as Newton's method to experimental or simulated kernels. The effect of these parameters may compensate for some of the approximations including: the monoenergetic model, and only single scattering events from the forward scatter model.

Along with the before-mentioned approximations of this model, there are a couple of limitations in calculating forward scatter using (\ref{equ:pre_forward}) instead of (\ref{equ:classic_for}). Firstly, it is calculated using the physical measurements $\boldsymbol{y}$, which will themselves be contaminated by the scatter one is attempting to estimate, and will lead to a biased estimate. In an attempt to mitigate this, scatter estimation may be performed iteratively, where $\boldsymbol{y}$ is first corrected for scatter, then used as the basis for the next estimate as in \citet{Sun2010}. This process then acts as a deconvolution, which is known to be non-exact with the presence of noise. This iterative correction also presents a second problem in cases when the scatter expectation is larger that the measurements themselves, which is possible with CBCT \citep{Siewerdsen2001}, and where the $\log(\cdot)$ will be undefined. Applying thresholding to the corrected $\boldsymbol{s}$ to ensure it remains positive as in \citet{Sun2010,Chang2014} will allow the calculation of (\ref{equ:pre_forward}), but at the expense of introducing a bias.

Due to the high dimensionality of the CBCT measurements, one can perform the convolution in (\ref{equ:simple_conv}) efficiently through use of the FFT. With this, the order of complexity can be reduced from $\mathcal{O}(N_\mathrm{proj}N_u^2N_v^2)$ to $\mathcal{O}(N_\mathrm{proj}N_uN_v\log(N_uN_v))$ for a detector with dimensions $N_u \times N_v$. The re-expression of (\ref{equ:simple_conv}) employing FFT operations is
\begin{equation}
\boldsymbol{s}\{k\}(u,v) = \mathcal{F}^{-1}\left[\mathcal{F}\left(\boldsymbol{p}\{k\}(u,v)\right)\odot\mathcal{F}\left(\boldsymbol{g}(u,v)\right)\right]  \mbox{ for }  i=1,...,N_\mathrm{proj},
\end{equation}
where $\mathcal{F}[\cdot]$ and $\mathcal{F}^{-1}[\cdot]$ are FFT and inverse FFT (IFFT) operations. To avoid aliasing and to allow an efficient FFT decomposition, some degree of zero padding is necessary, which may typically double the effective number of samples.

It was demonstrated in \citet{Sun2010} that even for homogeneous slabs of water, both the parameters of the forward scatter calculation $K$, $h_1$ and $h_2$, and the width of the Gaussian kernels $c_\mathcal{N}$ and $c_\mathcal{B}$ are dependent on the thickness of the slab for a typical energy spectrum. To account for this and still enable fast convolution, they decomposed the projection into several thickness groups, each with a different set of parameters. On top of this, they also include factor to account for inhomogeneities at edges, and linearly modulate the kernel amplitude with the object thickness to approximate more realistic scatter media. In combination, this is known as the fast adaptive SKS (fASKS) and can be expressed as \citep{Sun2010}
\begin{align} \label{equ:fasks}
\boldsymbol{s}\{k\}(u,v) = &(1-\gamma\boldsymbol{\tau}\{k\}(u,v))\odot\mathcal{F}^{-1}\left[\sum_{j}^{N_\mathcal{R}}\mathcal{F}\left(\boldsymbol{\mathcal{R}}_j\{k\}(u,v)\odot\boldsymbol{c}_j\{k\}(u,v)\odot\boldsymbol{p}_j\{k\}(u,v)\right)\odot\mathcal{F}\left(\boldsymbol{g}_j(u,v)\right)\right] \nonumber \\
&+\gamma\mathcal{F}^{-1}\left[\sum_{j}^{N_\mathcal{R}}\mathcal{F}\left(\boldsymbol{\tau}\{k\}(u,v)\odot\boldsymbol{\mathcal{R}}_j\{k\}(u,v)\odot\boldsymbol{c}_j\{k\}(u,v)\odot\boldsymbol{p}_j\{k\}(u,v)\right)\odot\mathcal{F}\left(\boldsymbol{g}_j(u,v)\right)\right]
\end{align}
where $N_\mathcal{R}$ are the number of thickness groups, $\boldsymbol{\mathcal{R}}_j\{k\}(u,v)$ are binary masking functions to isolate the projection elements belonging to the appropriate thickness interval, $\boldsymbol{c}_j\{k\}(u,v)$ are factors to linearly attenuate the scatter kernels close to edges, $\gamma$ is a scalar constant to weight the strength of the thickness modulation, and $\boldsymbol{\tau}\{k\}(u,v)$ is the water equivalent thickness of the projection given as
\begin{equation}
\boldsymbol{\tau}\{k\}(u,v) = -\frac{1}{m_w}\log\left(\boldsymbol{y}\{k\}(u,v)\oslash\boldsymbol{I}_\mathrm{in}\{k\}(u,v)\right) \mbox{ for }  i=1,...,N_\mathrm{proj},
\end{equation}
where $m_w$ is the mass attenuation coefficient of water at some nominal photon energy, and $\oslash$ denotes the element-wise division. To reduce the bias associated in calculating (\ref{equ:pre_forward}), (\ref{equ:fasks}) is iteratively applied to the measurements four times in \citet{Sun2010}.

\section{Methodology} \label{sec:scat_method}
\subsection{Polyenergetic Scatter Modelling} \label{sec4:poly_scat_model}

\subsubsection{Polyenergetic Forward Scatter Derivation}
The derivation of the forward scatter factor for SKS for preprocessed estimation in Section~\ref{sec:conv_classical} made the assumption that the incident beam was monoenergetic, and was in terms of just the linearized attenuation coefficient $\mu$. However, one can derive this in a more accurate polyenergetic fashion, where the proportion of scattered X-ray at the point $D$ in Figure~\ref{fig:scatter_deriv} at a given energy $\xi$ is
\begin{equation}
I_D(\xi) = I_U(\xi)K(\xi)\rho_e(U)\exp\left(-\int_U^D \boldsymbol{\mu}(\xi,\ell) \, d\ell \right).
\end{equation}
There are a couple of differences here from (\ref{equ:scat_prob}). These are the explicit dependence on source energy $\xi$, and the dependence on the energy independent electron density $\rho_e$. The reason for this is that the attenuation through scattering events may be expressed in terms of the associated interactive cross sections as \citep{Jackson1981}
\begin{equation}
\mu_\mathrm{scatter}(\xi) = \rho N_g(\sigma_\mathrm{incoh}(\xi)+\sigma_\mathrm{coh}(\xi)),
\end{equation}
where $\sigma_\mathrm{incoh}(\xi)$ is incoherent Compton scatter, and $\sigma_\mathrm{coh}(\xi)$ are coherent scatter events. From the definition of relative electron density in (\ref{equ:red}) the proportionality to relative electron density can be seen. Furthermore, one can find the scalar constant $K(\xi)$ is related to the physics as
\begin{equation}
K(\xi) \propto \frac{\sigma_\mathrm{incoh}(\xi)+\sigma_\mathrm{coh}(\xi)}{\rho_\mathrm{water}N_{g,\mathrm{water}}},
\end{equation}
with the constant of proportionality being related to the size and position of the detector element. With this, having the scatter intensity proportional to $\mu$ as in (\ref{equ:scat_prob}) can be seen as an approximation and will implicitly include the non-scattering photoelectric interactions.

Similarly to the monoenergetic derivation in (\ref{equ:for_deriv_mono}), the forward scatter factor can be calculated by integrating all the individual scatter factors from $O$ to $D$, which can be expressed as
\begin{align} \label{equ:for_deriv_poly}
p(\xi) &= \int_O^D I_O(\xi)\exp\left(-\int_O^{U'} \boldsymbol{\mu}(\xi,\ell) \, d\ell \right)K(\xi)\rho_e(U')\exp\left(-\int_{U'}^D \boldsymbol{\mu}(\xi,\ell) \, d\ell \right) \,dU' \nonumber \\
&= I_O(\xi)\exp\left(-\int_O^D \boldsymbol{\mu}(\xi,\ell) \, d\ell \right) K(\xi) \int_O^D\boldsymbol{\rho}_e(\ell) \, d\ell,
\end{align}
where the combination of the two attenuation integrals can only be made since photons do not change energy with scattering events with no deflection \citep{Davisson1952,Jackson1981}. Since in practice one will have a finite detector size, there will be some small change in photon energy from Compton scatter, but the approximation in (\ref{equ:for_deriv_poly}) is still reasonable since the angles will be small. In any case, the energy dependency of the scatter factor $K(\xi)$ is retained to allow any compensation for this, despite the zero angle scattering not having a dependency on energy \citep{Klein1929}.

As in (\ref{equ:classic_for}), one can express this in a discretized setting for all measurements as
\begin{equation} \label{equ:forward_scatter}
p_i(\xi) = K(\xi)b_i(\xi)\exp(-[\boldsymbol{\Phi\mu}(\xi)]_i)[\boldsymbol{\Phi\rho}_e]_i   \mbox{ for }  i=1,...,N_\mathrm{ray}.
\end{equation}

A convenient property of (\ref{equ:forward_scatter}) when used within Polyquant reconstruction is that the energy dependent attenuation $\boldsymbol{\mu}(\xi)$ and the electron density $\boldsymbol{\rho}_e$ are readily available at each iteration without need for further computation. With this, given that scatter kernels are consistent within a given energy, then the same scatter model will be able to be used for any spectrum, given that is it known.

As with the conventional SKS derivation, (\ref{equ:forward_scatter}) does ignore multiple scattering that rejoins the line--of--sight path, and we will analyse in the next section how well this assumption holds.

\subsubsection{Polyenergetic Water Kernel Analysis}
\begin{figure}[!htb]
	\centering
	\begin{subfigure}[b]{0.45\textwidth}
		\includegraphics[width=\textwidth]{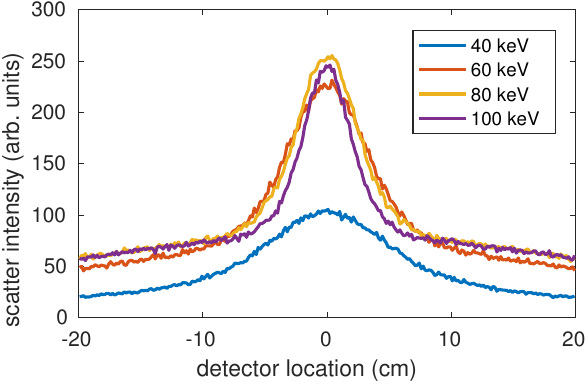}
		\caption{10 cm water slab kernels for different energies}
		\label{subfig:scat_kern1}
	\end{subfigure}
	\hfil
	\begin{subfigure}[b]{0.45\textwidth}
		\includegraphics[width=\textwidth]{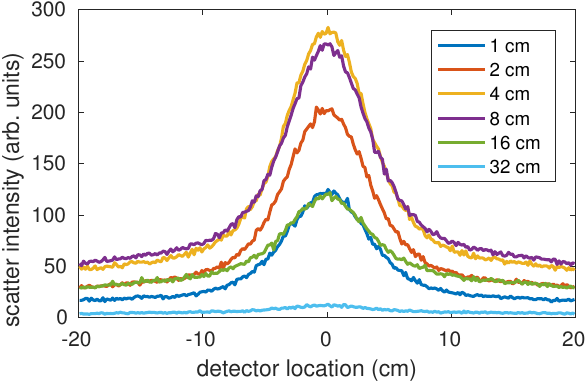}
		\caption{60 keV scatter kernel for different slab thickness}
		\label{subfig:scat_kern2}
	\end{subfigure}
	\caption{Examples of water slab scatter kernels from simulation plotted as detection intensity after detector response function and ADC (arb. units) against detector location (cm) along the $u$ dimension as shown in Figure~\ref{fig:scatter_deriv}.}
	\label{fig:scat_kernels}
\end{figure}

To build a convolutional model to utilize the polyenergetic scatter decomposition in (\ref{equ:forward_scatter}), we must investigate the nature of kernels from different energy sources. For this, we used the Monte Carlo particle simulation toolbox Gate \citep{Jan2011}, to calculate scatter profiles through slabs water of varying thickness but uniform cross sectional size of $80\times 80$ cm. We positioned a slab 50 cm from the planar detector and 100 cm from the point source. Some examples of these kernels are shown in Figure~\ref{fig:scat_kernels}.

Initial observations from Figure~\ref{fig:scat_kernels} are that there is indeed a strong link between the shape of the scatter profile and energy of the photon, and it appears that the width of the kernel decreases with an increase in energy as in Figure~\ref{subfig:scat_kern1}. As expected, there is also a strong link between thickness and kernel shape as shown in Figure~\ref{subfig:scat_kern2}, though it appears the width is roughly preserved. This suggests that instead of using a multi thickness decomposition for different kernel shapes as in (\ref{equ:fasks}), one could instead decompose in terms of energy. This would fit with the polyenergetic forward model derived in (\ref{equ:forward_scatter}).

\begin{figure}[!htb]
	\centering
	\begin{subfigure}[b]{0.45\textwidth}
		\includegraphics[width=\textwidth]{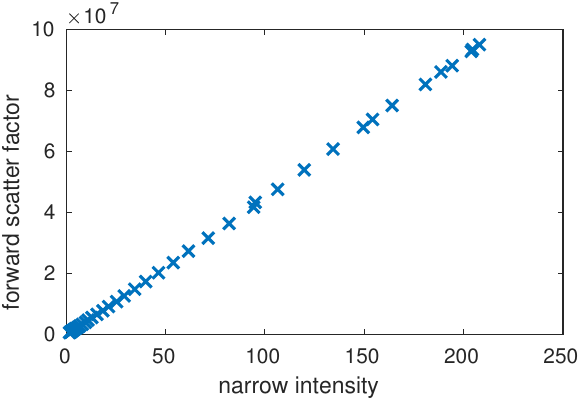}
		\caption{Analytic scatter factor against narrow amplitude}
		\label{subfig:narrow_fit}
	\end{subfigure}
	\hfil
	\begin{subfigure}[b]{0.45\textwidth}
		\includegraphics[width=\textwidth]{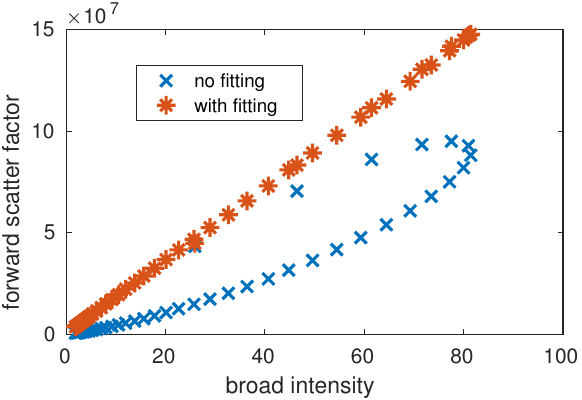}
		\caption{Analytic scatter factor against broad amplitude}
		\label{subfig:broad_fit}
	\end{subfigure}
	\caption{Relationship between narrow analytic forward scatter factors in (\ref{equ:forward_scatter}) and measured kernel amplitudes on water scatter data of 1--40 cm at 60 keV.}
	\label{fig:fit_scat_for}
\end{figure}

We have also plotted the relationship between the amplitude of the measured scatter kernels and the polyenergetic scatter factor with $K=1$ as calculated in (\ref{equ:forward_scatter}), which are shown in Figure~\ref{fig:fit_scat_for}.

It is clear from Figure~\ref{subfig:narrow_fit}, that our analytic model fits extremely well to the narrow Gaussian, where one can perform fitting by simply finding an appropriate proportionality constant $K(\xi)$. On the other hand, it does not agree entirely with the broad Gaussian as shown in Figure~\ref{subfig:broad_fit}.

However, by employing the same adjustment parameters $h_1,h_2$ as in classical scatter factor calculation in (\ref{equ:pre_forward}) \citep{Ohnesorge1999}, we are able to obtain a very good linear fit also, which is shown in Figure~\ref{fig:fit_scat_for}b. This then takes the form
\begin{equation}
p_{\mathcal{B},i}(\xi) = K(\xi)b_i(\xi)\exp(-[\boldsymbol{\Phi\mu}(\xi)]_i)^{h_1}[\boldsymbol{\Phi\rho}_e]_i^{h_2} \mbox{ for }  i=1,...,N_\mathrm{ray}.
\end{equation}
Due to properties of exponentials, this may be rewritten as
\begin{equation}
p_{\mathcal{B},i}(\xi) = K(\xi)b_i(\xi)\exp(-h_1[\boldsymbol{\Phi\mu}(\xi)]_i+h_2\log[\boldsymbol{\Phi\rho}_e]_i) \mbox{ for }  i=1,...,N_\mathrm{ray},
\end{equation}
where the parameters become scaling factors, lending itself to faster computational implementation. The fact that our derived single scatter forward model fits so well to the narrow Gaussian amplitude may be indicative that these result from single scatter events, whilst the broad signal is attributed to multiple scatters. However, through analysing the scattering order from our simulations, we have found this not to hold entirely, with single scatter also contributing significantly to the broad peak.

With these and other observations, we note that double Gaussian kernels from various energies and thickness of slab have the following properties:
\begin{itemize}
	\item The double Gaussian decomposition fits well for each energy and water thickness.
	\item The narrow field has a width that is dependent on the energy of the incident photons, and an amplitude that is directly proportional to the polyenergetic forward scatter factor in (\ref{equ:forward_scatter}).
	\item The broad field has a width which is independent of photon energy, and is proportional to the polyenergetic forward scatter factor in (\ref{equ:forward_scatter}) with appropriate fitting parameters $h_1,h_2$.
	\item With the above factors taken into account, there appears no additional dependence on thickness, suggesting the polyenergetic decomposition takes account for the change in shape observed in \citep{Sun2010}.
\end{itemize}
These observations are key to the implementation details of our scatter model given in the next section.

\subsubsection{Basic Polyenergetic Convolutional Model}
Using the polyenergetic forward model in (\ref{equ:forward_scatter}) and separate narrow and broad components, a basic form of double Gaussian polyenergetic scatter kernel accurate for a static slab may be written as 
\begin{align}
\boldsymbol{s}\{k\}(u,v) = \sum_{j=1}^{N_\xi}&\boldsymbol{p}_\mathcal{N}\{k\}(u,v,\xi_j)*\boldsymbol{g}_\mathcal{N}(u,v,\xi_j)+\\&\boldsymbol{p}_\mathcal{B}\{k\}(u,v,\xi_j)*\boldsymbol{g}_\mathcal{B}(u,v,\xi_j).
\end{align}
Since we have found that the broad field is roughly independent of energy, this may be simplified to
\begin{equation}
\boldsymbol{s}_\mathcal{B}\{k\}(u,v) = \left(\sum_{j=1}^{N_\xi}\boldsymbol{p}_\mathcal{B}\{k\}(u,v,\xi_j)\right)*\boldsymbol{g}_\mathcal{B}(u,v),
\end{equation}
which requires only a single convolution.

As with the preprocessed SKS, the convolution operations may be replaced by point-wise multiplication in the spatial frequency domain, whereby the calculation of scatter may be made as follows
\begin{equation}
\boldsymbol{s}\{k\}(u,v) = \mathcal{F}^{-1}\left[\boldsymbol{S}_\mathcal{N}\{k\}(\omega_u,\omega_v) + \boldsymbol{S}_\mathcal{B}\{k\}(\omega_u,\omega_v)\right] \mbox{ for }  i=1,...,N_\mathrm{proj},
\end{equation}
where $(\omega_u,\omega_v)$ are the spatial frequencies along the dimensions of the detector plane $(u,v)$, and
\begin{equation} \label{equ4:narrow_scat_fast}
\boldsymbol{S}_\mathcal{N}\{k\}(\omega_u,\omega_v) = \sum_{j=1}^{N_\xi}\mathcal{F}\left[\boldsymbol{p}_\mathcal{N}\{k\}(u,v,\xi_j)\right]\odot\mathcal{F}\left[\boldsymbol{g}_\mathcal{N}(u,v,\xi_j)\right],
\end{equation}
and
\begin{equation} \label{equ4:broad_scat_fast}
\boldsymbol{S}_\mathcal{B}\{k\}(\omega_u,\omega_v) = \mathcal{F}\left[\sum_{j=1}^{N_\xi}\boldsymbol{p}_\mathcal{B}\{k\}(u,v,\xi_j)\right]\odot\mathcal{F}\left[\boldsymbol{g}_\mathcal{B}(u,v)\right].
\end{equation}
We note that with this, the computational cost of a full scatter estimate involves $N_\mathrm{proj}$ IFFT operations and $N_\mathrm{proj}(N_\xi+1)$ FFT operations, since the Fourier transform of the Gaussian kernels need not be explicitly computed.

\subsection{Accounting for Detector Distance}
We already highlighted the fact that the forward scatter scalar constant $K(\xi)$ will in practice depend on the distance to the detector due to a finite detector size. On top of this, the width of the kernels will also vary as the distance to detector is varied, due to a magnification effect. In traditional preprocessing scatter calculation such as fASKS, these factors cannot be easily accounted for, since one does not know the position of the object. In our proposed integrated model however, this information is available from the current iterate, and may be exploited to further enhance the estimation accuracy.

\begin{figure}[!htb]
	\centering
	\includegraphics[width=0.5 \textwidth]{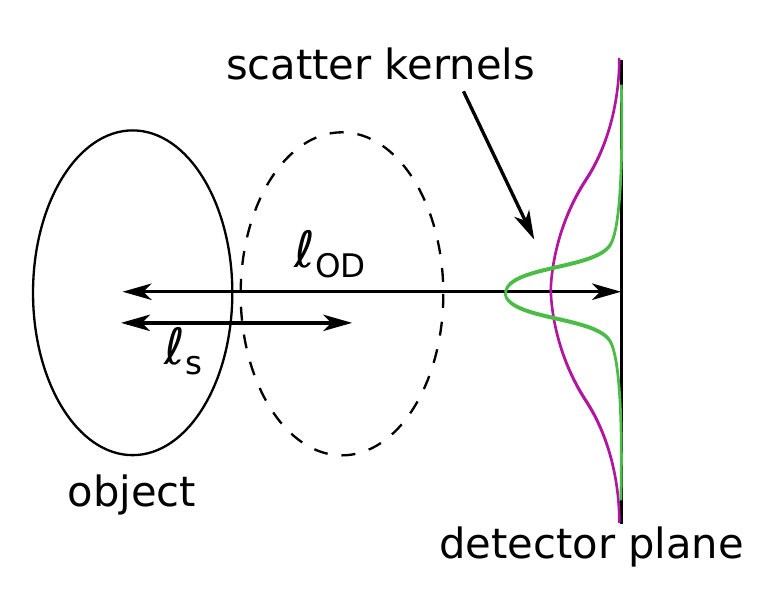}
	\caption{Scatter shift diagram illustrating magnification effect assuming the scattered photons originate from the center of the specimen}
	\label{fig:scatter_shift}
\end{figure}

With reference to Figure~\ref{fig:scatter_shift}, we can derive these correction factors based on the relative location of the object's center of mass. This makes the assumption that the mean scatter originates from the center of mass, which implies independence on the position of the source. As the object moves closer to the detector, we expect a decrease in Gaussian width, and an increase in amplitude due to the increase in area covered by the central pixel. By introducing a magnification factor $\zeta$, defined as
\begin{equation}
\zeta = \frac{\ell_{OD}-\ell_s}{\ell_{OD}},
\end{equation}
one would expect the width to change by a factor of $\zeta$, and the amplitude by a factor of $1/\zeta^2$. In practice, this shift distance $\ell_s$ will vary depending on the projection angle due to the change in scatter direction relative to the fixed coordinate system of the specimen. This is compensated rotating the coordinate system with
\begin{equation}
\ell_s = x_0 \cos\theta - y_0 \sin\theta,
\end{equation}
where $(x_0,y_0)$ are the coordinates at the center of the specimen and $\theta$ is the angle of the gantry.

\begin{figure}[!htb]
	\centering
	\begin{subfigure}[b]{0.45\textwidth}
		\includegraphics[width=\textwidth]{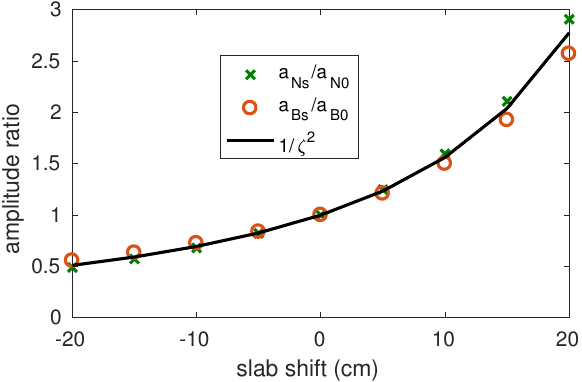}
		\caption{Shift amplitude factors}
		\label{subfig:mag_amp}
	\end{subfigure}
	\hfil
	\begin{subfigure}[b]{0.45\textwidth}
		\includegraphics[width=\textwidth]{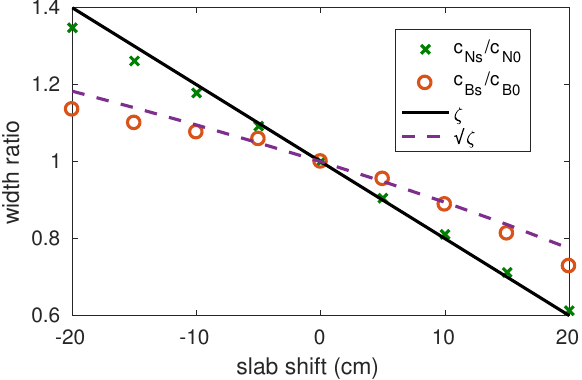}
		\caption{Shift width factors}
		\label{subfig:mag_shift}
	\end{subfigure}
	\caption{Shift amplitude and width factors for 10 cm water slab with 60 keV source.}
	\label{fig:mag_factors}
\end{figure}

To test if this model holds, we evaluated the change in scatter width and forward scatter factors for the narrow and broad components. The experiment consisted of the Monte Carlo simulation of a 10 cm slab of water shifted axially from its original position 50 cm away from the detector. We have plotted the resulting factors in Figure~\ref{fig:mag_factors} for a 60 keV source, and have found the same trend holds throughout other energies.

It can be observed from Figure~\ref{fig:mag_factors} that the amplitudes of both a narrow and broad Gaussian scatter components indeed fit $1/\zeta^2$ very well. The narrow Gaussian width factor also follows the expected value very well, though the broad factor appears to have a weaker dependence on shifts in the object. We note, however, that it empirically fits well to a modified factor of $\sqrt{\zeta}$, which we adopt in practice.

\subsection{Compensation for Object Edges} \label{sec:edge_effects}
A critical effect from inhomogeneous objects is the reduction of the scatter magnitude towards edges, as observed in \citet{Sun2010,Maltz2008,Ruhrnschopf2011b}. This occurs since photons scattered away from an edge into a lower attenuating material such as air are far less likely to undergo further scattering events. Using a slab model such as SKS at the edges leads to an overestimation of scatter behind the thicker parts of the object, which will typically have the highest scatter to primary ratio. The consequences of this are not only that the attenuation through the line--of--sight measurements will be overestimated, but photon starvation and streak artifacts are likely to occur in the reconstruction \citep{Chang2014} after its compensation.

In the fASKS work of \citet{Sun2010}, the authors observe an exponential drop in kernel magnitude towards an edge, which they approximate with the linear weighting factor $\boldsymbol{c}\{k\}(u,v)$ that can be applied between interfaces of thickness groups. 

We employ a similar mechanism for use in our polyenergetic scatter model that instead produces a continuous modification of the kernel amplitude. From simulations on water slabs and elliptical tubes, we have found that the reduction in kernel amplitude is only significant on the broad Gaussian, which takes the following exponential form
\begin{equation} \label{equ:edge_factor}
\tilde{\boldsymbol{p}}_\mathcal{B} = \boldsymbol{p}_\mathcal{B}\exp\left(-\frac{\boldsymbol{t}_u^2+\boldsymbol{t}_v^2}{c_\mathcal{B}^2}\right),
\end{equation}
where $\boldsymbol{t}_u$ and $\boldsymbol{t}_v$ are continuous edge weights in the detector plane coordinates, and $\tilde{\boldsymbol{p}}_\mathcal{B}$ is the compensated broad forward scatter factor.

\begin{figure}[!htb]
	\centering
	\includegraphics[width=0.5 \textwidth]{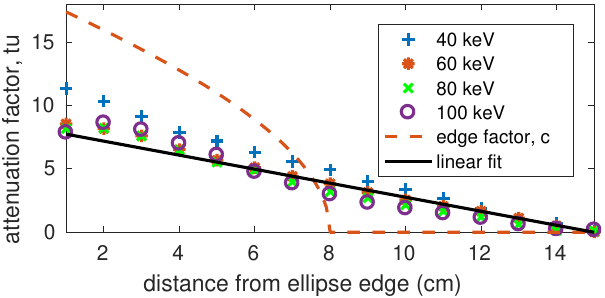}
	\caption{Broad kernel scatter shifts from simulated elliptical tube object fitted through nonlinear least squares at various source energies.}
	\label{fig:edge_factor}
\end{figure}

In Figure~\ref{fig:edge_factor}, we have shown the least squares fitting for $\boldsymbol{t}_u$ according to the model in (\ref{equ:edge_factor}). The data was generated from Monte Carlo simulations of an elliptical tube with semi-major axis of 15 cm semi-minor axis of 7.5 cm and length of 80 cm, where in each run the object was laterally shifted relative to the detector plane axis $u$. We are interested in the elliptical tube object since it crudely approximates the shape of a human body lying on a couch. With this, we also plotted the factor $\boldsymbol{c}\{k\}(u,v)$ as given in \citet{Sun2010}, though within the exponential model of (\ref{equ:edge_factor}).

From the nature of the kernel factors as observed in Figure~\ref{fig:edge_factor}, the increasing kernel attenuation factor towards the edge of the elliptical object can be seen. While this trend is somewhat caught by the linear edge factor, $\boldsymbol{c}\{k\}(u,v)$ from \citep{Sun2010}, it does not capture the right functional form. Other observations from Figure~\ref{fig:edge_factor} are that there is not a strong dependence on the incident energy, that there is a very small deviation from the slab model at the center of the object, and the trend in $\boldsymbol{t}_u$ is approximately linear.

Motivated by this linear trend, we opt to use the following edge weighting function
\begin{equation} \label{equ:shift_factor}
\boldsymbol{t}_u\{k\}(u,v) = k_\mathrm{edge}\boldsymbol{\tau}_e\{k\}(u,v)\frac{\partial\boldsymbol{\tau}_e\{k\}(u,v)}{\partial u},
\end{equation}
where $\boldsymbol{\tau}_e$ is the electron density projection as $\boldsymbol{\tau}_e = \boldsymbol{\Phi}\boldsymbol{\rho}_e$, and $\frac{\partial\boldsymbol{\tau}_e\{k\}(u,v)}{\partial u}$ is the spatial derivative of the electron density projection that can be calculated in practice as a discrete gradient. The factor $k_\mathrm{edge}$ is a scalar parameter to set the strength of the edge compensation effect. For isotropy, the factor $\boldsymbol{t}_v\{k\}(u,v)$ is calculated in the same way but with the derivative with respect to $v$. 

\begin{figure}[!htb]
	\centering
	\includegraphics[width=0.35 \textwidth]{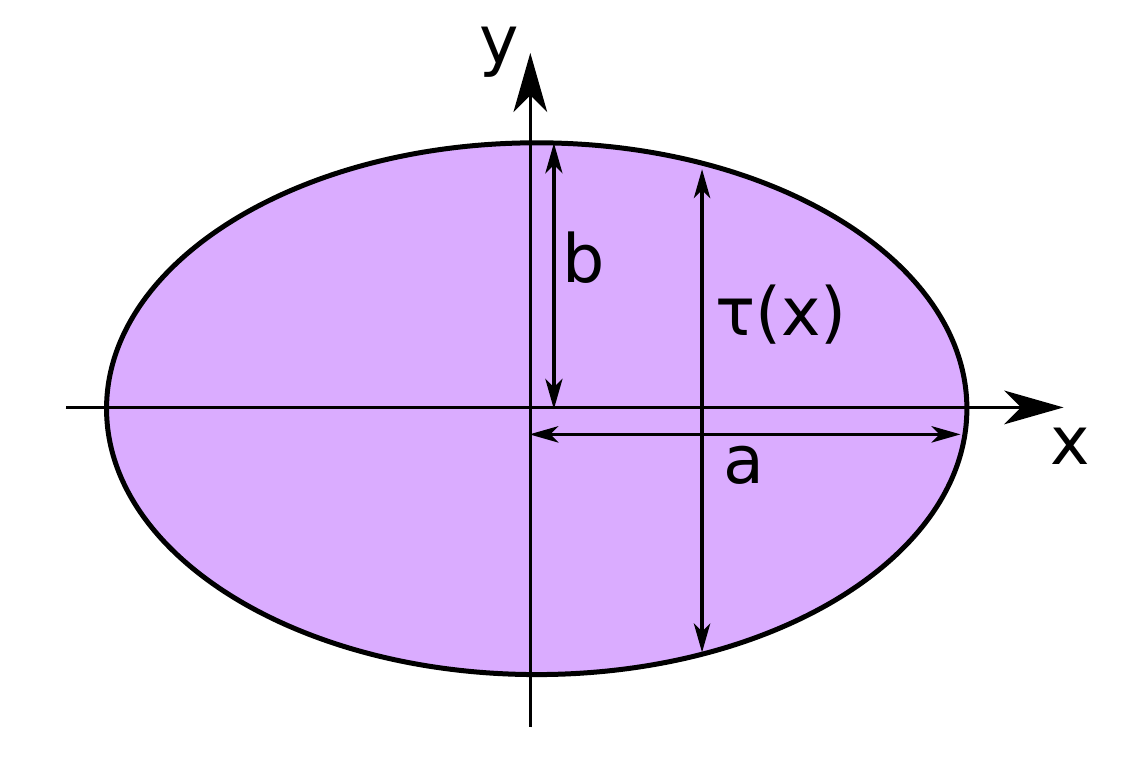}
	\caption{Diagram of ellipse for edge factor derivation.}
	\label{fig:ellipse_digram}
\end{figure}

It is easy to show that for a simple ellipse model, the scatter estimate in (\ref{equ:shift_factor}) gives the correct linear trend observed in Figure~\ref{fig:edge_factor}. Figure~\ref{fig:ellipse_digram} shows an ellipse with major and minor semi axes lengths of $a$ and $b$ on $x$ and $y$ respectively. The thickness at a given point along $x$ is denoted by $\tau(x)$. It can be shown that the variation in thickness with respect to $x$ is expressed as
\begin{equation}
\frac{d \tau(x)}{d x} = -\frac{b^2x}{a^2\tau(x)}.
\end{equation}
Using the form of (\ref{equ:shift_factor}) in this case becomes
\begin{equation}
k_\mathrm{edge}\tau(x)\frac{d \tau(x)}{d x} \propto \frac{b^2}{a^2}x,
\end{equation}
which is now linear with the distance towards the edge, and dependent on the squared ratio between height and width of the ellipse. This second feature is also consistent with the experimental observation in \citet{Sun2010}, where they find an increase in edge effect with thicker objects.

To provide robustness of this method to rapidly varying projections, we found smoothing the electron density thickness $\boldsymbol{\tau}_e\{k\}(u,v)$ is beneficial. In the experimental sections, we employ a 2D Gaussian filter with standard deviation 1.5 cm. This smoothing also allows the edge compensation to be active towards sharp edges.

\subsection{Complete Polyenergetic Scatter Estimation Model and Fitting}
Combining the various elements, we note our complete scatter model --- PolySKS --- may be formed using the factors
\begin{equation} \label{equ:narrow_for}
\boldsymbol{S}_\mathcal{N}\{k\}(\omega_u,\omega_v) = \sum_{j=1}^{N_\xi}\mathcal{F}\left[\frac{1}{\zeta^2}\boldsymbol{p}_\mathcal{N}\{k\}(u,v,\xi_j)\right]\odot\mathcal{F}\left[\exp\left(-\frac{\boldsymbol{u}^2+\boldsymbol{v}^2}{\zeta^2c_\mathcal{N}^2(\xi_j)}\right)\right],
\end{equation}
and
\begin{align} \label{equ:broad_for}
\boldsymbol{S}_\mathcal{B}\{k\}(\omega_u,\omega_v) = &\mathcal{F}\left[\frac{1}{\zeta^2}\sum_{j=1}^{N_\xi}\boldsymbol{p}_\mathcal{B}\{k\}(u,v,\xi_j)\exp\left(-\frac{\boldsymbol{t}_u^2+\boldsymbol{t}_v^2}{c_\mathcal{B}^2}\right)\right] \nonumber\\
\odot&\mathcal{F}\left[\exp\left(-\frac{\boldsymbol{u}^2+\boldsymbol{v}^2}{\zeta c_\mathcal{B}^2}\right)\right].
\end{align}
and as before the estimate is then
\begin{equation}
\boldsymbol{s}\{k\}(u,v) = \mathcal{F}^{-1}\left[\boldsymbol{S}_\mathcal{N}\{k\}(\omega_u,\omega_v) + \boldsymbol{S}_\mathcal{B}\{k\}(\omega_u,\omega_v)\right] \mbox{ for }  i=1,...,N_\mathrm{proj},.
\end{equation}

In calculating (\ref{equ:narrow_for}), the exact form of forward scatter factor used is
\begin{equation} \label{equ:fast_scat_fac}
\boldsymbol{p}_\mathcal{N}\{k\}(u,v,\xi_j) = K_\mathcal{N}(\xi_j)\boldsymbol{b}\{k\}(u,v,\xi_j)\odot\exp(-[\boldsymbol{\Phi\mu}(\xi_j)]\{k\})\odot[\boldsymbol{\Phi\rho}_e]\{k\},
\end{equation}
whilst the broad component of (\ref{equ:broad_for}) utilizes the fitting parameters as
\begin{equation}
\boldsymbol{p}_\mathcal{B}\{k\}(u,v,\xi_j) = K_\mathcal{B}(\xi_j)\boldsymbol{b}\{k\}(u,v,\xi_j)\odot\exp(-h_1(\xi_j)[\boldsymbol{\Phi\mu}(\xi_j)]\{k\}+h_2(\xi_j)\log([\boldsymbol{\Phi\rho}_e]\{k\})).
\end{equation}

In order to use this model, one requires to find the energy dependent parameters $c_\mathrm{N}$, $K_\mathcal{N}$, $K_\mathcal{B}$, $h_1$ and $h_2$, as well as the energy independent parameters $c_\mathrm{B}$ and $k_\mathrm{shift}$ from (\ref{equ:shift_factor}) to calculate $\boldsymbol{t}_u$ and $\boldsymbol{t}_v$. Apart from the edge factor parameters, these were fitted to data from simulating scatter through water slabs having a thickness range of 1 cm to 40 cm with 1 cm increment, and for a number of different photon energies. In each case, we used a detector response function derived from the energy absorption coefficient of Cesium Iodine (CsI) with a thickness of 0.6 mm to replicate a photo absorption scintillator, and a detector of $40\times 40$ cm with resolution $256\times 256$. We then minimized the nonlinear least squares error against this data to our model using Newton's method, assuming a fixed broad width of 35 cm. These parameters are shown in Table~\ref{tab:parameters}.

\begin{table}[!htb]
	\caption{PolySKS parameters from water kernel fitting used throughout the experimental section.}
	\label{tab:parameters}
	\centering
	$
	\begin{array}{c|c|c|c|c|c||c|c|c}
	\text{energy (keV)} & K_\mathcal{N} & c_\mathcal{N} \text{(cm)} & K_\mathcal{B} & h_1 & h_2 & c_\mathcal{B} \text{(cm)} & k_\mathrm{edge} \text{(full fan)} &  k_\mathrm{edge} \text{(half fan)}\\
	\hline
	40 & 1.38\times 10^{-6} & 6.91 & 3.51\times 10^{-7} & 0.876 & 1.10 & 35 & 2.35 & 1.57 \\
	60 & 1.40\times 10^{-6} & 4.77 & 3.16\times 10^{-7} & 0.828 & 1.15 & - & - & - \\
	80 & 1.39\times 10^{-6} & 3.62 & 2.97\times 10^{-7} & 0.804 & 1.20 & - & - & - \\
	100 & 1.40\times 10^{-6} & 2.90 & 2.79\times 10^{-7} & 0.791 & 1.25 & - & - & - \\
	120 & 1.40\times 10^{-6} & 2.44 & 2.66\times 10^{-7} & 0.785 & 1.29 & - & - & - \\
	
	\hline
	\end{array}
	$
\end{table}

From Table~\ref{tab:parameters}, one can see typical parameters from our fitting. It is interesting to see that although we included the energy dependence on $K_\mathcal{N}$ to account for finite detector effects, there is almost no dependence from our fitting, which agrees with theory from \citet{Klein1929}. Also interesting is the significant deviation of the narrow width and broad forward model parameters throughout energy. A critical factor to consider is that the scatter scaling factors $K_\mathcal{N}$ and $K_\mathcal{B}$ will be dependent on the area of a detector element, and they should be appropriately scaled against our detector area of 0.0244 cm$^2$ for other resolutions.

For the perturbation factor $k_\mathrm{edge}$, we calculated the average factor across energies for the elliptical tube experiment presented in Figure~\ref{fig:edge_factor} according to the model in (\ref{equ:shift_factor}). When we repeated this for an offset detector, we found a reduction of around 33\%. This may be due to wider cone angles of the geometry having less of an effect from edges perpendicular to the detector. For the other parameters however, we found they hold in both full fan or half fan acquisitions.

\section{Incorporating the Polyenergetic Scatter Model in Polyquant}
Although the polyenergetic model as derived in Section~\ref{sec:scat_method} is a general model for scatter that may be used in various algorithms or applications, its exact formulation works particularly well in the Polyquant framework, Polyquant--PolySKS. To derive an algorithm, we start with the gradient term of the negative log-likelihood NLL function with respect to the electron density, which has the parameter
\begin{equation} \label{equ4:polysks_diff}
\boldsymbol{d}(\boldsymbol{\rho}_e) = \boldsymbol{y}\oslash\left(\sum_{j=1}^{N_\xi}\boldsymbol{\psi}(\boldsymbol{\rho}_e,\xi_j)+\boldsymbol{s}(\boldsymbol{\rho}_e)\right) - \mathbf{1},
\end{equation}
where $\boldsymbol{s}(\boldsymbol{\rho}_e)$ is the updated scatter estimate based on the current iterate using (\ref{equ:fast_scat_fac}), and $\boldsymbol{\psi}(\boldsymbol{\rho}_e,\xi_j)$ is defined in (\ref{equ:psi}). This is then used to calculate the derivative as
\begin{equation} \label{equ4:polysks_deriv}
\frac{\partial\mathrm{NLL}(\boldsymbol{\rho}_e;\boldsymbol{y})}{\partial\boldsymbol{\rho}_e} \approx
\sum_{i=1}^{N_f}\boldsymbol{f}_i(\boldsymbol{\rho}_e) 
\odot\boldsymbol{\Phi}^T
\left[\sum_{j=1}^{N_\xi}\alpha_i(\xi_j)\boldsymbol{\psi}(\boldsymbol{\rho}_e,\xi_j)\odot\boldsymbol{d}(\boldsymbol{\rho}_e)\right],
\end{equation}
in the same manner as the regular Polyquant method \cite{Mason2017b}. We note a subtle difference is the dependence of the scatter term with $\boldsymbol{\rho}_e$. With this, the exact formulation of the gradient would include a term from the derivative of the scatter w.r.t. the current estimate, which for convolutional models would be an extra filtering term. We have not included this in (\ref{equ4:polysks_deriv}), since we have found it only has a weak dependence on the NLL and hence can be ignored. For these reasons, we have ignored this dependence and have symbolized this non-exactness by the `$\approx$' sign.

Calculating the term $\boldsymbol{s}(\boldsymbol{\rho}_e)$ requires from (\ref{equ:forward_scatter}) the terms $\boldsymbol{\Phi \mu}(\xi)$ for each energy and $\boldsymbol{\Phi \rho}_e$. For the first term, we note from the Polyquant formulation in (\ref{equ:gen_fit}) we have
\begin{equation}
\boldsymbol{\Phi} \hat{\boldsymbol{\mu}}(\xi_j) = \sum_{i=1}^{N_f}\alpha_i(\xi_j)\boldsymbol{\Phi}[\boldsymbol{f}_i(\boldsymbol{\rho}_e)\odot\boldsymbol{\rho}_e]+\beta_i(\xi_j)\boldsymbol{\Phi}\boldsymbol{f}_i(\boldsymbol{\rho}_e),
\end{equation}
where the expensive factors to compute $\boldsymbol{\Phi}[\boldsymbol{f}_i(\boldsymbol{\rho}_e)\odot\boldsymbol{\rho}_e]$ and $\boldsymbol{f}_i(\boldsymbol{\rho}_e)$ are already available at each iteration. Additionally, due to the properties of the threshold functions given in (\ref{equ:threshold}), we can find that
\begin{equation}
\boldsymbol{\Phi}\boldsymbol{\rho}_e = \sum_{i=1}^{N_f}\boldsymbol{\Phi}[\boldsymbol{f}_i(\boldsymbol{\rho}_e)\odot\boldsymbol{\rho}_e],
\end{equation}
which also only requires projections that are already available at each iteration anyway.

\subsection{Fast Scatter Fused Reconstruction Algorithm}
From the new formulation with the fusion of a convolutional scatter model in (\ref{equ4:polysks_deriv}) with (\ref{equ4:polysks_diff}), Polyquant--PolySKS, can be treated with the same Prox-Polyquant algorithm as given in \citet{Mason2017b}. In order to increase the efficiency of the algorithm --- important for large CBCT data sets --- we consider here an accelerated variant of the original Polyquant. For this, we employ the concept of combining `ordered subsets' with momentum as in \citet{Kim2015a,Wang2015}, given in Algorithm~\ref{alg2}. We highlight that this is essentially an implementation of FISTA \citep{Beck2009a} applied to our objective function, though that each calculation of a gradient operates on some subset $\mathcal{S}^k$ of the full NLL.

\begin{algorithm}                      
	\caption{Fast-Polyquant}
	\label{alg2}
	\begin{algorithmic}                    
		\REQUIRE Subset selection scheme for $\mathcal{S}^k$, regularization constant $\lambda$, Lipschitz approximation $L_0$, number of iterations $N_\mathrm{iter}$, number of subset divisions $N_\mathrm{split}$, and starting point $\boldsymbol{\rho}_e^1$.
		\STATE $\boldsymbol{\rho}_e^1 \leftarrow \mathbf{1}$
		\STATE $\boldsymbol{z}^0 \leftarrow \boldsymbol{\rho}_e^1$
		\STATE $\delta \leftarrow 1.9N_\mathrm{split}/L_0$
		\FOR{$k = 1,2,\ldots,N_\mathrm{iter}$}
		\STATE $\boldsymbol{z}^{k} \leftarrow \mathbf{prox}_{\delta\lambda R}\left(\boldsymbol{\rho}_e^{k}-\delta\frac{\partial\mathrm{NLL}(\boldsymbol{\rho}_e^{k};\boldsymbol{y}\{\mathcal{S}^k\},\mathcal{S}^k)}{\partial\boldsymbol{\rho}_e}\right)$
		\STATE $t^{k+1} \leftarrow \frac{1+\sqrt{1+4(t^k)^2}}{2}$
		\STATE $\boldsymbol{\rho}_e^{k+1} \leftarrow \boldsymbol{z}^{k}+\left(\frac{t^k-1}{t^{k+1}}\right)(\boldsymbol{z}^{k}-\boldsymbol{z}^{k-1})$
		\ENDFOR
		\RETURN $\boldsymbol{\rho}_e^{N_\mathrm{iter}}$
	\end{algorithmic}
\end{algorithm}

In Algorithm~\ref{alg2}, each iteration uses a gradient calculated on only a subset of the NLL, defined as
\begin{equation}
\mathrm{NLL}(\boldsymbol{x}^{k};\boldsymbol{y},\mathcal{S}^k) = \sum_{i\in\mathcal{S}^k}\sum_{j=1}^{N_\xi}\psi_i(\boldsymbol{x},\xi_j)+s_i - y_i\log\left(\sum_{j=1}^{N_\xi}\psi_i(\boldsymbol{x},\xi_j)+s_i\right),
\end{equation}
given some subset $\mathcal{S}^k\subset[1,\dots,N_\mathrm{ray}]$, where the projection indices array is defined as
\begin{equation}
\boldsymbol{\iota}^k = [l^k:N_\mathrm{split}:(N_\mathrm{proj}-N_\mathrm{split}+l^k)],
\end{equation}
where $N_\mathrm{split}$ is the number of subset divisions, and $\boldsymbol{\iota}$ is a subset index array that selects a subset as
\begin{equation}
\hat{\boldsymbol{y}}_{j} = [\boldsymbol{y}\{\boldsymbol{\iota}^k\}]_j \mbox{ for } j = 1,\dots,\frac{N_\mathrm{proj}}{N_\mathrm{split}},
\end{equation}
with $\hat{\boldsymbol{y}}$ as the subset of the full set of measurements $\boldsymbol{y}$. A simple consecutive ordered selection of the subsets can be performed as $l^k = (k\mod N_\mathrm{split})+1$. Selection of $l^k$ has been shown to have impact of the robustness of these types of approaches \citep{Kim2015a}, for which we have adopted the `bit reversal ordering' of \citet{Herman1993}. This has the effect of reducing the correlation and error accumulation between consecutive iterations \citep{Kim2015a}. The number of iterations $N_\mathrm{iter}$ is set to a multiple of the subset divisions such that $N_\mathrm{iter} = N_\mathrm{epoch}N_\mathrm{split}$, where $N_\mathrm{epoch}$ is the number of effective data passes (epochs), which is selected to allow for empirical convergence.

The term $\mathbf{prox}_{\delta R}$ is the proximity operator defined as
\begin{equation} \label{equ:prox}
\mathbf{prox}_{\delta R}(\boldsymbol{w}) = \argmin_{\boldsymbol{\rho}_e\in\mathcal{C}} \frac{1}{2} \|\boldsymbol{w}-\boldsymbol{\rho}_e\|_2^2 + \delta\lambda R(\boldsymbol{\rho}_e),
\end{equation}
where $\boldsymbol{w}$ is a temporary input, $R(\cdot)$ is a convex regularization function, $\lambda$ is the regularization constant, and $\mathcal{C}$ is a constraint set on electron density. In practice, we use a box constraint so that $0\leq \rho_{e,i}\leq \zeta \mbox{ for }  i=1,...,N_\mathrm{vox.}$, where $\zeta$ is the maximum allowable density value (possibly $\zeta=\infty$), and the constraint set ensures non-negative density values. Without loss of generality we opt for TV regularization for $R(\cdot)$. By using TV, which promotes piecewise smoothness --- something expected with homogeneous slabs of biological tissue --- the term (\ref{equ:prox}) may be solved as outlined in \citet{Beck2009a} for example.

As with the original `Prox-Polyquant' algorithm in \citet{Mason2017b}, we use an estimate of the Lipschitz constant $L_0$, computed as
\begin{equation}
L_0 = \left\|\boldsymbol{\Phi}^T\left[\left(\sum_{j=1}^{N_\xi}\alpha_1^2(\xi_j)\boldsymbol{b}(\xi_j)\right)\odot\boldsymbol{\Phi1}\right]\right\|_\infty,
\end{equation}
where $\|\cdot\|_\infty$ is the infinity norm that selects the maximum of the diagonal of the Hessian at point $\boldsymbol{0}$.

\citet{Kim2015a} also introduced a second variant of acceleration using the algorithm of \citet{Nesterov2005} that provided better performance in \citet{Wang2015}. However, we found the FISTA variant performed better with our model and with TV regularization. 

\section{Evaluation from Simulation} \label{sec4:eval_simulation}
In the first experiment, we evaluate the accuracy of our proposed integrated scatter model, Polyquant--PolySKS, with comparison to precomputed scatter. To synthesize the data, we performed a Monte Carlo simulation of a CBCT system using the software Gate \citep{Jan2011} according to the method outlined in \cite{Mason2017c}.

We generated two acquisitions of the female adult reference computational phantom (ARCP) \citep{icrp2009} from the pelvis and head region, using half and full fan scans respectively. In both cases, we used a planar detector of $40\times 30$ cm, placed 150 cm from the source and 50 cm from the center of rotation. For the half fan pelvis case, we offset the detector by 16 cm and used a shaped source to replicate the effect of a bow-tie filter. We also used a appropriate source profile for the full fan head case. For both scans, we simulated a total of $1\times 10^{11}$ photons over 160 projections and with a detector resolution of $256\times 128$. The image resolutions were those of the original ARCP at $299\times 137$ and 60 slices of 0.484 cm thickness. We do not include any scatter grid in our simulation.

All methods under test use the Polyquant reconstruction technique \citep{Mason2017b}, with identical TV regularization, allowing a fair comparison between scatter estimation approaches. We will thereby refer to the reconstruction method Polyquant--PolySKS only by the scatter estimation method PolySKS. The spectra used were equivalent to 100 kVp and 120 kVp tube potentials for the head and pelvis cases respectively, which we equally sampled into 21 energies. The spectra were generated using the SpekCalc software \citep{Poludniowski2009c} for a tungsten anode at $30^\circ$. The system operators $\boldsymbol{\Phi}$ and $\boldsymbol{\Phi}^T$ were implemented as separable footprint projections \citep{Balter2011} within the Michigan Image Reconstruction Toolbox \citep{Fessler} in \textsc{Matlab}. The methods we will use for comparison are:
\begin{itemize}
	\item \textit{Scatter free}: we have entirely removed all scatter from the numerical data, which represents the gold standard of any scatter mitigation method.
	\item \textit{No correction}: scatter is included in the measurements, but no attempt is made to correct it, to show the extent of scatter induced artifacts and represents the `worst case'.
	\item \textit{Pre-fASKS}: Polyquant with pre-calculated fASKS calculated according to (\ref{equ:fasks}), applied iteratively to the measurements 10 times. For fairness, we refit all parameters to the same simulated water slabs as were used to inform the PolySKS parameters in Table~\ref{tab:parameters}. For the perturbation factors, we used the same edge factor as presented in the original publication \citep{Sun2010}, and used $\gamma=0.04$ and $\gamma=0.03$ for the head and pelvis cases respectively, selected to give the best fitting against our elliptical shift data used to fit $k_\mathrm{edge}$.
	\item \textit{Int-fASKS}: an identical scatter model and parameters as in \textit{Pre-fASKS}, but performed at each iteration using the forward scatter factor (\ref{equ:classic_for}) in place of its approximation in (\ref{equ:pre_forward}). This does not exploit either the spectral information nor the electron density.
	\item \textit{PolySKS}: our proposed polyenergetic integrated scatter model as detailed in Section~\ref{sec4:poly_scat_model} to give Polyquant--PolySKS, according to parameters given in Table~\ref{tab:parameters}.
\end{itemize}
Additionally, we provide a comparison to a hardware based estimation approach using a beam stop array. which is detailed in Section~\ref{sec:bsa}. For each method, we performed 10 epochs in the head case and 20 epochs for the pelvis. It seemed the lower amount of attenuating material in the head case meant these methods convergence significantly faster.

\begin{figure}[!htb]
	\centering
	\begin{subfigure}[b]{0.19\textwidth}
		\includegraphics[trim=2.5cm 0.5cm 2.5cm 0cm,clip=true,width=\textwidth]{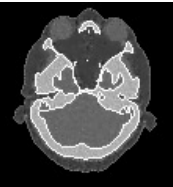}
		\caption{Scatter free}
	\end{subfigure}
	\begin{subfigure}[b]{0.19\textwidth}
		\includegraphics[trim=2.5cm 0.5cm 2.5cm 0cm,clip=true,width=\textwidth]{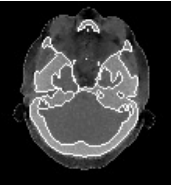}
		\caption{No correction}
	\end{subfigure}
	\begin{subfigure}[b]{0.19\textwidth}
		\includegraphics[trim=2.5cm 0.5cm 2.5cm 0cm,clip=true,width=\textwidth]{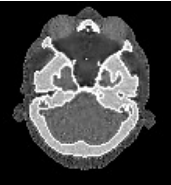}
		\caption{Pre-fASKS}
	\end{subfigure}
	\begin{subfigure}[b]{0.19\textwidth}
		\includegraphics[trim=2.5cm 0.5cm 2.5cm 0cm,clip=true,width=\textwidth]{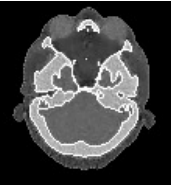}
		\caption{Int-fASKS}
	\end{subfigure}
	\begin{subfigure}[b]{0.19\textwidth}
		\includegraphics[trim=2.5cm 0.5cm 2.5cm 0cm,clip=true,width=\textwidth]{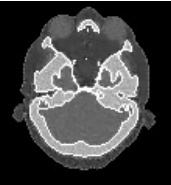}
		\caption{PolySKS}
	\end{subfigure}
	\begin{subfigure}[b]{0.19\textwidth}
		\includegraphics[trim=2.5cm 0.5cm 2.5cm 0cm,clip=true,width=\textwidth]{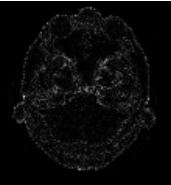}
		\caption{Scatter free error}
	\end{subfigure}
	\begin{subfigure}[b]{0.19\textwidth}
		\includegraphics[trim=2.5cm 0.5cm 2.5cm 0cm,clip=true,width=\textwidth]{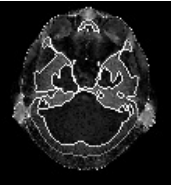}
		\caption{No estimate error}
	\end{subfigure}
	\begin{subfigure}[b]{0.19\textwidth}
		\includegraphics[trim=2.5cm 0.5cm 2.5cm 0cm,clip=true,width=\textwidth]{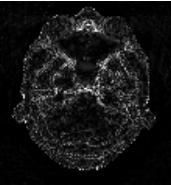}
		\caption{Pre-fASKS error}
	\end{subfigure}
	\begin{subfigure}[b]{0.19\textwidth}
		\includegraphics[trim=2.5cm 0.5cm 2.5cm 0cm,clip=true,width=\textwidth]{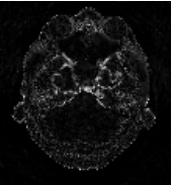}
		\caption{Int-fASKS}
	\end{subfigure}
	\begin{subfigure}[b]{0.19\textwidth}
		\includegraphics[trim=2.5cm 0.5cm 2.5cm 0cm,clip=true,width=\textwidth]{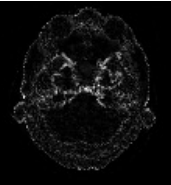}
		\caption{PolySKS error}
	\end{subfigure}
	\caption{Visual results of slice 36 from the head CBCT data with display window [0.5,1.4] to aid visualization of soft tissue and reconstruction artifacts; absolute errors have display window [0,0.3].}
	\label{fig4:head_sim_recon}
\end{figure}

\begin{figure}[!htb]
	\centering
	\begin{subfigure}[b]{0.32\textwidth}
		\includegraphics[trim=1cm 0cm 1cm 0.5cm,clip=true,width=\textwidth]{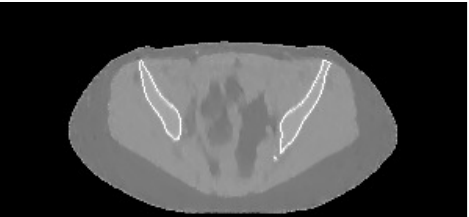}
		\caption{Scatter free, slice 43}
	\end{subfigure}
	\begin{subfigure}[b]{0.32\textwidth}
		\includegraphics[trim=1cm 0cm 1cm 0.5cm,clip=true,width=\textwidth]{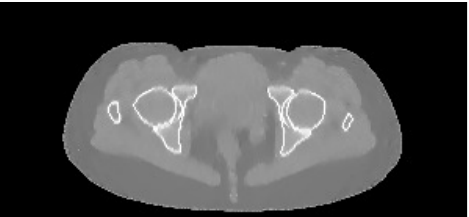}
		\caption{Scatter free, slice 30}
	\end{subfigure}
	\begin{subfigure}[b]{0.32\textwidth}
		\includegraphics[trim=1cm 0cm 1cm 0.5cm,clip=true,width=\textwidth]{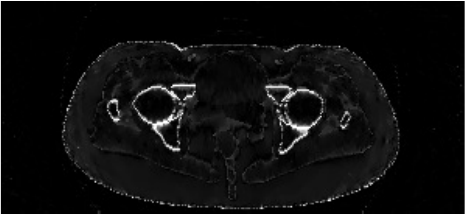}
		\caption{Scatter free error, slice 30}
	\end{subfigure}
	\begin{subfigure}[b]{0.32\textwidth}
		\includegraphics[trim=1cm 0cm 1cm 0.5cm,clip=true,width=\textwidth]{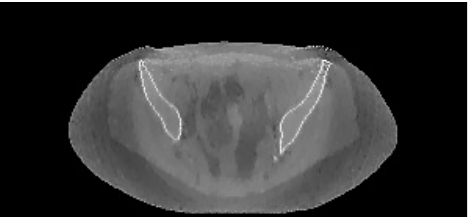}
		\caption{No correction, slice 43}
	\end{subfigure}
	\begin{subfigure}[b]{0.32\textwidth}
		\includegraphics[trim=1cm 0cm 1cm 0.5cm,clip=true,width=\textwidth]{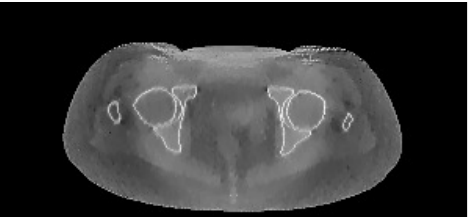}
		\caption{No correction, slice 30}
	\end{subfigure}
	\begin{subfigure}[b]{0.32\textwidth}
		\includegraphics[trim=1cm 0cm 1cm 0.5cm,clip=true,width=\textwidth]{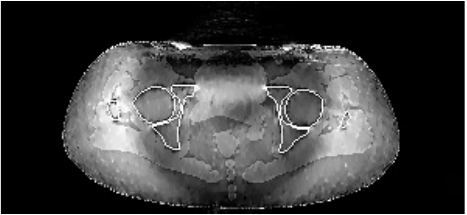}
		\caption{No correction error, slice 30}
	\end{subfigure}
	\begin{subfigure}[b]{0.32\textwidth}
		\includegraphics[trim=1cm 0cm 1cm 0.5cm,clip=true,width=\textwidth]{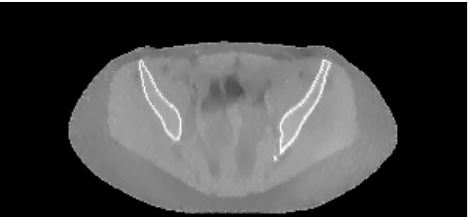}
		\caption{Pre-fASKS, slice 43}
	\end{subfigure}
	\begin{subfigure}[b]{0.32\textwidth}
		\includegraphics[trim=1cm 0cm 1cm 0.5cm,clip=true,width=\textwidth]{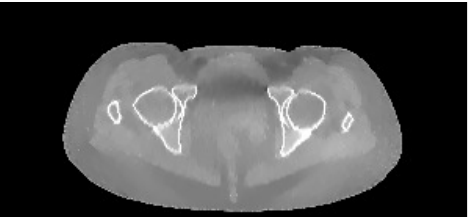}
		\caption{Pre-fASKS, slice 30}
	\end{subfigure}
	\begin{subfigure}[b]{0.32\textwidth}
		\includegraphics[trim=1cm 0cm 1cm 0.5cm,clip=true,width=\textwidth]{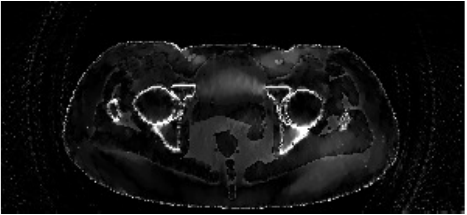}
		\caption{Pre-fASKS error, slice 30}
	\end{subfigure}
	\begin{subfigure}[b]{0.32\textwidth}
		\includegraphics[trim=1cm 0cm 1cm 0.5cm,clip=true,width=\textwidth]{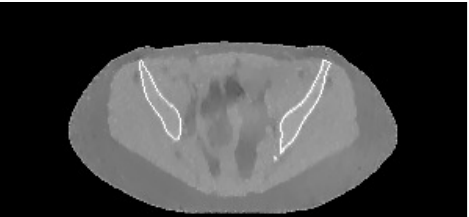}
		\caption{Int-fASKS, slice 43}
	\end{subfigure}
	\begin{subfigure}[b]{0.32\textwidth}
		\includegraphics[trim=1cm 0cm 1cm 0.5cm,clip=true,width=\textwidth]{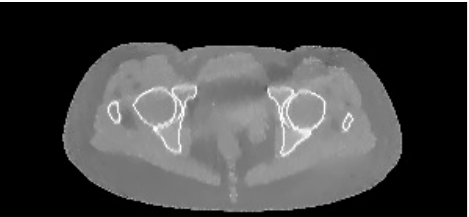}
		\caption{Int-fASKS, slice 30}
	\end{subfigure}
	\begin{subfigure}[b]{0.32\textwidth}
		\includegraphics[trim=1cm 0cm 1cm 0.5cm,clip=true,width=\textwidth]{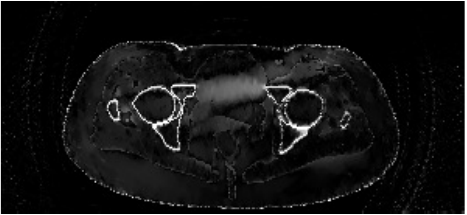}
		\caption{Int-fASKS error, slice 30}
	\end{subfigure}
	\begin{subfigure}[b]{0.32\textwidth}
		\includegraphics[trim=1cm 0cm 1cm 0.5cm,clip=true,width=\textwidth]{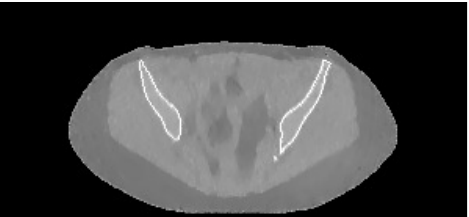}
		\caption{PolySKS, slice 43}
	\end{subfigure}
	\begin{subfigure}[b]{0.32\textwidth}
		\includegraphics[trim=1cm 0cm 1cm 0.5cm,clip=true,width=\textwidth]{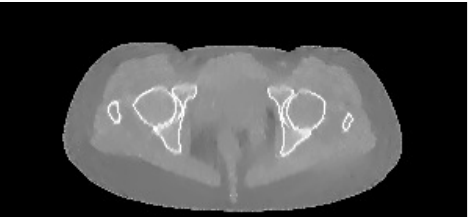}
		\caption{PolySKS, slice 30}
	\end{subfigure}
	\begin{subfigure}[b]{0.32\textwidth}
		\includegraphics[trim=1cm 0cm 1cm 0.5cm,clip=true,width=\textwidth]{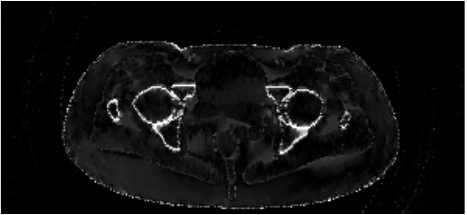}
		\caption{PolySKS error, slice 30}
	\end{subfigure}
	\caption{Visual results of slices 43 and 30 from the pelvis CBCT data with display window [0.5,1.4] to aid visualization of soft tissue and reconstruction artifacts; absolute errors have display window [0,0.3].}
	\label{fig4:pelvis_sim_recon}
\end{figure}

To evaluate the performance of the methods, we have shown reconstructed images in Figures~\ref{fig4:head_sim_recon} and \ref{fig4:pelvis_sim_recon}. We also quantified the accuracy through the root--mean--squared--error (RMSE) through slices 18--42 --- corresponding to slices in the cone angle --- and these are tabulated in Table~\ref{tab4:sim_results}. The various slices selected for display were those exhibiting significant structure or artifacts. The relative quantitative performance of the methods under test was found to be consistent throughout all well sampled slices. 

For the scatter contaminated head phantom reconstruction, shown illustratively in Figure~\ref{fig4:head_sim_recon}b and \ref{fig4:head_sim_recon}g, the scatter leads to visible shading artifacts, especially around the edges and close to the bones, as well as a significant underestimation in the bony structures. Of the scatter correction methods, both the integrated scatter estimation approaches Int-fASKS and PolySKS perform similarly, whilst the Pre-fASKS still exhibits shading artifacts and has a higher level of noise. Quantitatively, from the second column in Table~\ref{tab4:sim_results}, this trend is replicated, with all correction methods considerably closing the gap between `scatter free' and `no estimate', with the PolySKS giving the highest accuracy. The difference between the fASKS methods is especially interesting as it implies one must incur some loss from the deconvolution and using (\ref{equ:pre_forward}) in place of (\ref{equ:classic_for}).

The pelvis reconstructions are then shown in Figure~\ref{fig4:pelvis_sim_recon}. Here, scatter is much more of a critical factor, with the `no estimate' in Figures~\ref{fig4:pelvis_sim_recon}d--f showing huge errors throughout the image. In this case, both Pre-fASKS and Int-fASKS exhibit strong shading artifacts also, especially in the region between the bones. Although PolySKS does have a number of overestimated regions in the fat regions visible in the error plot in Figures~\ref{fig4:pelvis_sim_recon}o, it is generally a good match to the `scatter free' reconstruction. Quantitatively from Table~\ref{tab4:sim_results}, the same trend as the head case can be seen, with the Int-fASKS giving roughly a 9\% drop in error over Pre-fASKS, and PolySKS giving another 9\% reduction over that.

\begin{table}[!htb]
	\caption{Quantitative results: relative electron density ($\rho_e$) RMSE of slices 18--42.}
	\label{tab4:sim_results}
	\centering
	$
	\begin{array}{c|c|c|c|c|c|c|}
	\cline{2-7}
	&\multicolumn{3}{c}{\text{Polyquant}}&\multicolumn{3}{|c|}{\text{FDK}}\\
	\cline{2-7}
	\text{Scheme} & \text{Head RMSE} & \text{Pelvis RMSE} & \text{Run time* (s)} & \text{Head RMSE} & \text{Pelvis RMSE} & \text{Run time (s)}\\
	\hline
	\text{Scatter free} & 0.0112 & 0.0322 & 150 & 0.0631 & 0.0683 & 1.2\\
	\text{No estimate} & 0.0413 & 0.122 & 150 & 0.0707 & 0.102 & 1.2 \\
	\text{BSA} & 0.0165 & 0.0430 & 150 & 0.0639 & 0.0695 & 1.5\\
	\text{Pre-fASKS} & 0.0258 & 0.0479 & 170 & 0.0686 & 0.0906 & 21\\
	\text{Int-fASKS} & 0.0183 & 0.0435 & 200 & - & - & -\\
	\text{PolySKS} & \mathbf{0.0152} & \mathbf{0.0398}& 240 & - & - & -\\
	\hline
    \multicolumn{6}{c}{\text{*Polyquant reconstruction times given for head case}}\\
	\end{array}
	$
\end{table}

For computational comparison, we have indicated the run times of the various approaches in Table~\ref{tab4:sim_results} for the same hardware. In this case, all methods are run in \textsc{Matlab} on a commodity PC with an 8 core Intel Xeon E5-1660 v3 processor running at 3 GHz with 32 GB of RAM. To show the cost of the iterative reconstruction, we have also included times for the Feldkamp--Davis--Kress (FDK) method \citep{Feldkamp1984}, for the preprocessing scatter estimation approaches. Although the cost of FDK is considerably lower than Polyquant, there is also a vast loss in accuracy. Between the Polyquant approaches, all are at the same order of computational time, but PolySKS does have a 60\% higher cost over no estimation in this case. Potential approaches for mitigating this additional cost are discussed in Section~\ref{comp:feas}.

\subsection{Beam Stop Array Comparison} \label{sec:bsa}
We have also quantitatively evaluated the relative performance of a hardware based scatter estimation technique, beam stop array (BSA). For our implementation of BSA, we replicated the experimental set up in \citet{Ning2004} through simulation. We positioned the array between source and specimen, and extracted the detected values behind their shadows as a sparse sampling of scatter. We then used smoothing and cubic B-spline interpolation to cover the entire detector. Unlike in \citet{Ning2004}, where the BSA was subsampled at a low dose, we used a separate fully sampled acquisition, to show a best case performance for this method. With this, the effective radiation dose increased by $\sim 98\%$ over other tested methods.

The performance of BSA is competitive, with a relative error of 8\% against our proposed PolySKS, and stands as the second best method under test, although this gap will likely broaden if evaluated at the same dose. A notable benefit of BSA is its rapid computational time, which is found to be a fraction of the cost of an FDK reconstruction. With this, a trade off between dose, computation and accuracy exists, and an appropriate choice will depend on the application. For maximum quantitative performance at a minimal dose, Polyquant--PolySKS is seen to be the strongest method under test.

\section{Validation from Real Data} \label{sec4:eval_realdata}
In this section, we tested our method on real data from CBCT scans of physical phantom objects. For this, we have two quantitative CIRS (Computerized Imaging Reference Systems, Inc., Norfolk, VA (USA)) phantoms with acquisitions on a Varian\textsuperscript{\textregistered} TrueBeam\texttrademark\ On-Board Imager\textsuperscript{\textregistered} (Varian Medical Systems, Inc., Palo Alto, CA (USA)).

In each of the two phantom studies we compare these three approaches:
\begin{itemize}
	\item \textit{No estimate}: Polyquant reconstruction with no scatter estimate. This is included to indicate the magnitude of the scatter artifacts and to represent a worst case.
	\item \textit{Pre-fASKS}: Polyquant reconstruction is performed using the estimate of scatter generated by the TrueBeam system as preprocessing. This is an instance of fASKS shown in (\ref{equ:fasks}), but with the manufacturers parameters, and with a constant tailed triangle modulation of the kernel shapes to account for the scatter grid on the scanner, the inclusion of a detector glare model \citep{Sun2010} and a heuristic to improve upon scatter inaccuracies from the couch as in \citep{Sun2011}.
	\item \textit{PolySKS}: our proposed integrated quantitative scatter model as outlined in Section~\ref{sec4:poly_scat_model}. In this case, we use the very same model and parameters from Table~\ref{tab:parameters} as used in the simulated section, though with the inclusion of a similar modulation for the presence of a scatter grid as with Pre-fASKS.
\end{itemize}

For every method, we ran Polyquant for 50 epochs. Unlike the simulated experiments, we did not evaluate the Int-fASKS approach since we did not have means to modify the system's scatter approach, and we have already demonstrated the superiority of PolySKS over this in the numerical results. Additionally, we were unable to evalutate BSA, since the hardware was unavailable for our system.

\subsection{CIRS Insert Phantom}
In the first case, we investigate reconstructions of a CIRS electron density insert phantom. The acquisition consisted of 896 projections over $360^\circ$ with a 100 kVp tube potential and a current of 20 mA over a 15 ms pulse duration.

The phantom itself has a cross sectional diameter of 18 cm and a thickness of 5 cm. This poses a challenge for scatter estimation due to its thinness, since it naturally deviates significantly from the underlying assumed slab model.  

Table~\ref{tab4:cirs_reference} shows material information for the tissue inserts in the phantom. Of particular interest is the relative electron densities $\rho_e$, since these will act as the ground truths against which to compare our quantitative estimates. To calculate the estimates from the reconstructions, we took the mean value from a circular region of radius 1 cm of each insert. Both the body of the phantom and the central left and right inserts are manufactured from `water equivalent' resin. Since we are pursuing the quantitatively accurate estimation of soft tissues, we have not evaluated these. However, since we would like to investigate the shading effect of scatter on homogeneous slabs, we do assess the deviation with the body structure.  

\begin{table}[!htb]
	\caption{Material definitions of insert phantom.}
	\label{tab4:cirs_reference}
	\centering
	$
	\begin{array}{c|c|c|c|c|c|c|}
	\text{Label} & \text{A} & \text{B} & \text{C} & \text{D} & \text{E} & \text{F}\\
	\hline
	\text{Material} & \text{Adipose} & \text{Muscle} & \text{Soft bone} & \text{Muscle} & \text{Breast} & \text{Liver} \\
	\text{Density (g/cm$^3$)}  & 0.96 & 1.06 & 1.16 & 1.06 & 0.99 & 1.07 \\
	\text{RED, } \rho_e & 0.949 & 1.043 & 1.117 & 1.043 & 0.949 & 1.052 \\
	\hline
	\end{array}
	$
\end{table}

\begin{figure}[!htb]
	\centering
	\begin{subfigure}[b]{0.32\textwidth}
		\includegraphics[trim=1cm 1cm 1cm 1cm,clip=true,width=\textwidth]{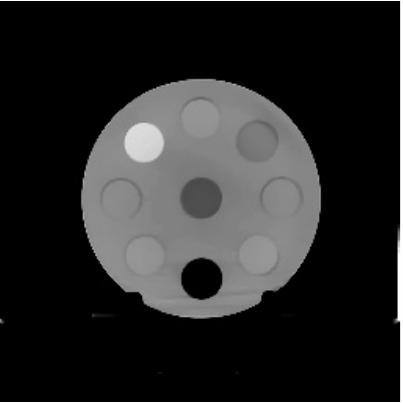}
		\caption{No estimate}
	\end{subfigure}
	\begin{subfigure}[b]{0.32\textwidth}
		\includegraphics[trim=1cm 1cm 1cm 1cm,clip=true,width=\textwidth]{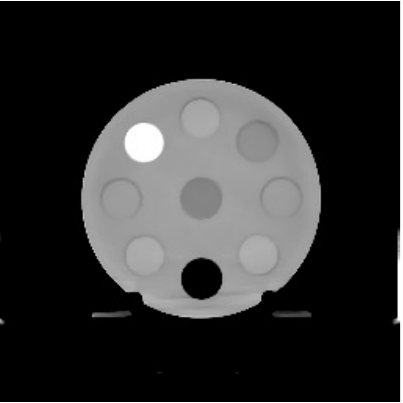}
		\caption{Pre-fASKS}
	\end{subfigure}
	\begin{subfigure}[b]{0.32\textwidth}
		\includegraphics[trim=1cm 1cm 1cm 1cm,clip=true,width=\textwidth]{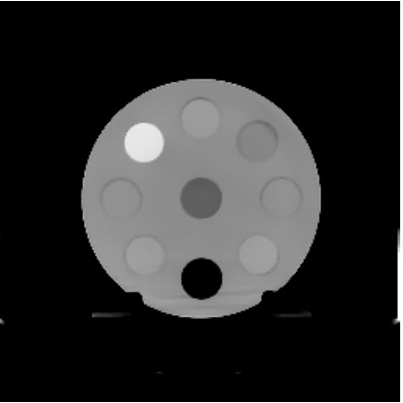}
		\caption{PolySKS}
	\end{subfigure}
	
	\caption{Illustrations of reconstructed volumes of insert phantom of central slice with display window [0.8,1.2].}
	\label{fig:cirs_insert_recon}
\end{figure}

\begin{figure}[!htb]
	\centering
	\begin{subfigure}[b]{0.32\textwidth}
		\includegraphics[trim=1cm 1cm 1cm 1cm,clip=true,width=\textwidth]{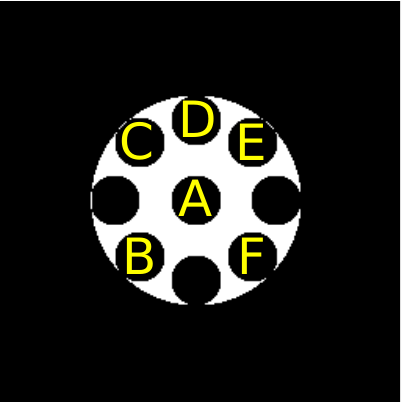}
		\caption{Regions of interest}
	\end{subfigure}
	\begin{subfigure}[b]{0.65\textwidth}
		\includegraphics[width=\textwidth]{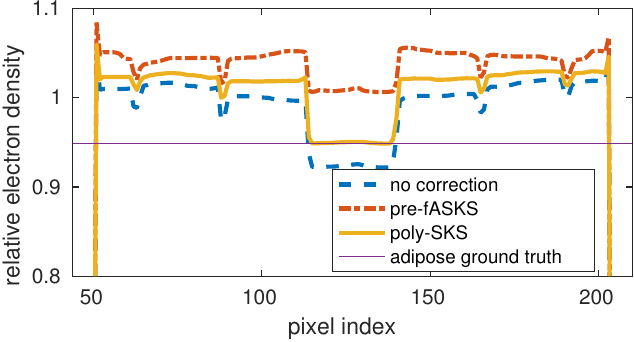}
		\caption{Profiles through center of reconstructions}
	\end{subfigure}
	\caption{Insert location labels, along with slab ROI mask, and line profiles through center of reconstructions.}
	\label{fig4:cirs_profile}
\end{figure}

Reconstructions from the insert phantom are shown in Figure~\ref{fig:cirs_insert_recon}. To supplement this, we have also plotted a line profile horizontally through their centers in Figure~\ref{fig4:cirs_profile}. From observation, all three methods produce similar reconstructions. In the case of no estimate in Figure~\ref{fig:cirs_insert_recon}a, there is a visible shading towards the center of the object also visible in its profile, which is very typical of a scatter artifact. Whilst both the Pre-fASKS and PolySKS compensate for this, PolySKS appears to generate more of a uniform slab estimation, with the former producing a raised estimate at the center and edges of the profile in Figure~\ref{fig4:cirs_profile}. This overestimation is likely caused by overestimating scatter, perhaps arising the thinness of the phantom.

\begin{table}[!htb]
	\caption{Quantitative results: relative electron density accuracy of insert phantom.}
	\label{tab:cirs_results}
	\centering
	$
	\begin{array}{c|c|c|c|c|c|c|c|c}
	\cline{2-8}
	&\multicolumn{6}{c}{\text{Tissue accuracy (\% error)}}&\multicolumn{1}{|c|}{\text{Slab deviation (std)}}\\
	\cline{2-8}
	\text{Scheme} & \text{A} & \text{B} & \text{C} & \text{D} & \text{E} & \text{F} & \text{ROI} & \text{Run time (min)}\\
	\hline
	\text{No estimate} & -2.73 & -1.08 & \mathbf{1.72} & -1.04 & \mathbf{1.70} & -1.75 & 7.74\times 10^{-3} & 78\\
	\text{pre-fASKS} & 6.23 & 2.32 & 8.01 & 2.14 & 5.07 & 1.69 & 5.38\times 10^{-3} & 78\text{*}\\
	\text{PolySKS} & \mathbf{0.0718} & \mathbf{0.390} & 3.91 & \mathbf{0.171} & 3.56 & \mathbf{-0.277} & \mathbf{4.77\times 10^{-3}} & 120\\
	\hline
	\multicolumn{9}{c}{\text{*computation time of fASKS unknown but likely small relative to cost of iterative reconstruction}}
	\end{array}
	$
\end{table}

Table~\ref{tab:cirs_results} then shows the quantitative analysis of the reconstructions. It is clear that Pre-fASKS produces an overestimation of the electron densities across the board, which in many cases represents a worse performance than from no estimate. PolySKS on the other hand performs very well in most cases, having an error within the stated material manufacturing tolerance of 1\%, with the exception of the breast and bone materials. Although it may appear disconcerting that the no estimate has the most accurate performance in the bone and breast cases, we note that these are still overestimations. Since in reality there will be some non-zero scatter behind these structures, and the addition of any scatter compensation will raise the estimation of density, it is unsurprising that both Pre-fASKS and PolySKS overestimate this further, so we suggest this is unlikely arising from a deficit in scatter estimation. Possible causes for the discrepancy may be inaccurate spectrum modelling or effects from objects outside the field of view such as the couch. 

The other reported value in Table~\ref{tab:cirs_results} is the slab deviation, which is the standard deviation of all pixels within the ROI in Figure~\ref{fig4:cirs_profile}a. Here, we note that either scatter modelling strategy offers a large increase in uniformity by at least 30\% over `no estimate', with a further 11\% increase from Pre-fASKS to PolySKS. This result is unsurprising considering the flatness of the profiles in Figure~\ref{fig4:cirs_profile}.

Due to the higher dimensionality of the data, the run times as shown in Table~\ref{tab:cirs_results} are significantly longer than those from the simulated cases in Table~\ref{tab:cirs_results}, but the relative additional computation of incorporating PolySKS is roughly preserved. 

In summary from this experiment: PolySKS results in the most uniformity throughout homogeneous structures, and is the only variant offering highly quantitatively electron density estimation, apart from the bone and breast that are overestimated in every case.

\subsection{STEEV Head Phantom}
In the second case, we performed reconstruction of the CIRS STEEV head phantom. In this case, the phantom consists of complex resin structures to mimic the attenuation and form of a human head. Again, we wish to analyse the quantitative accuracy and uniformity of the homogeneous materials. The acquisition used the same source settings as the insert phantom, and with 501 projections over a $200^\circ$ gantry rotation. 

\begin{figure}[!htb]
	\centering
	\begin{subfigure}[b]{0.45\textwidth}
		\includegraphics[trim=1cm 2cm 1cm 0cm,clip=true,width=\textwidth]{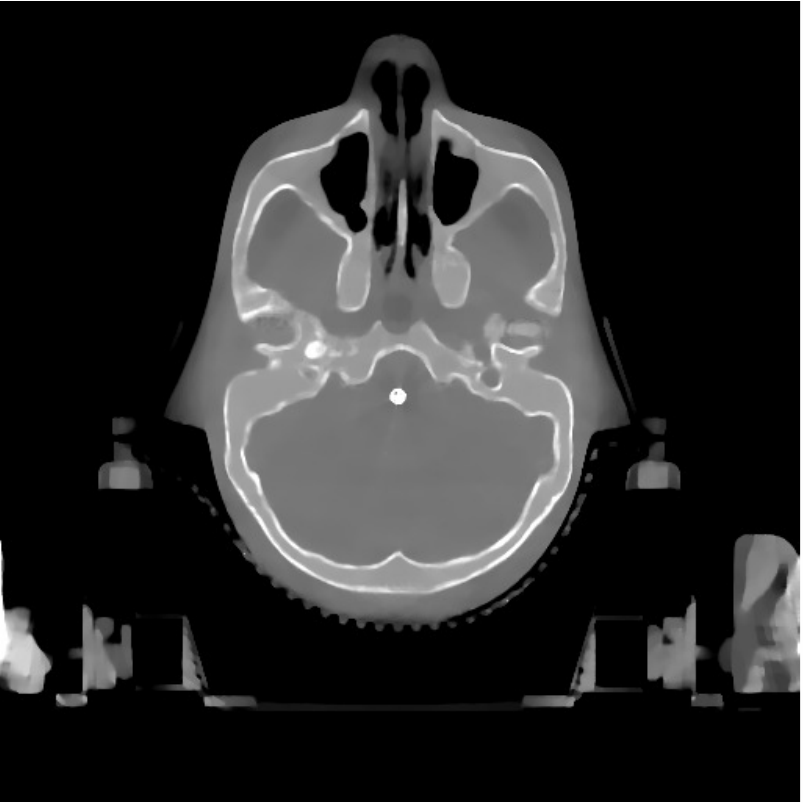}
		\caption{No estimate}
	\end{subfigure}
	\begin{subfigure}[b]{0.45\textwidth}
		\includegraphics[trim=1cm 2cm 1cm 0cm,clip=true,width=\textwidth]{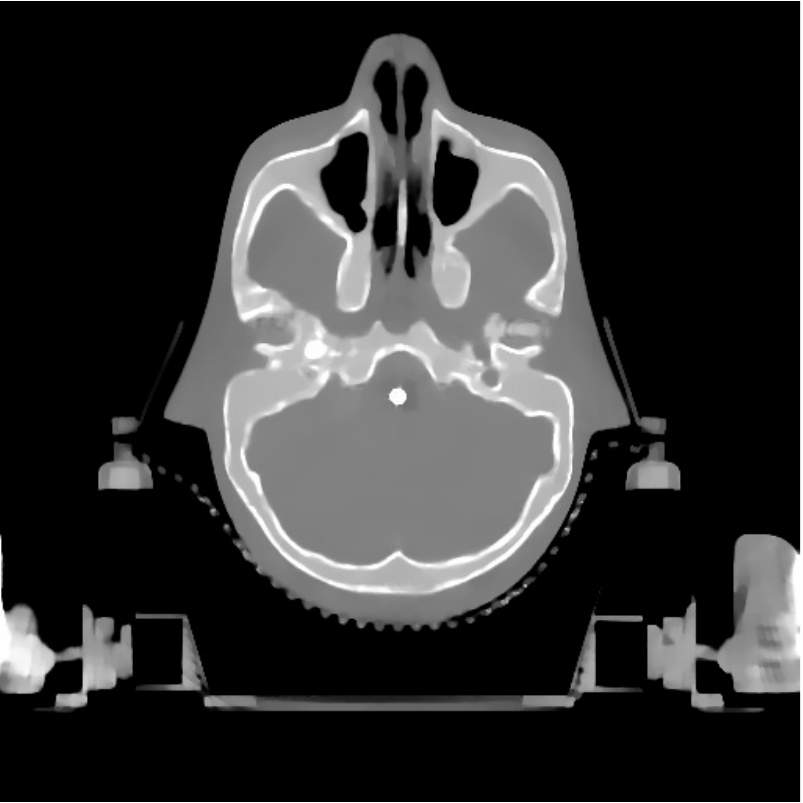}
		\caption{Pre-fASKS}
	\end{subfigure}
	\begin{subfigure}[b]{0.45\textwidth}
		\includegraphics[trim=1cm 2cm 1cm 0cm,clip=true,width=\textwidth]{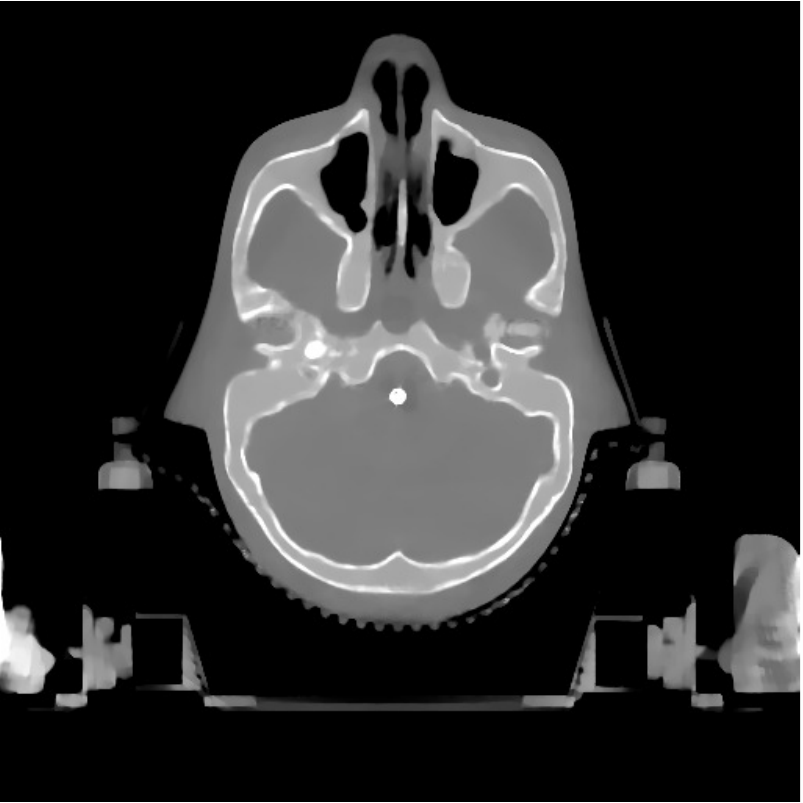}
		\caption{PolySKS}
	\end{subfigure}
	\begin{subfigure}[b]{0.45\textwidth}
		\includegraphics[trim=1cm 2cm 1cm 0cm,clip=true,width=\textwidth]{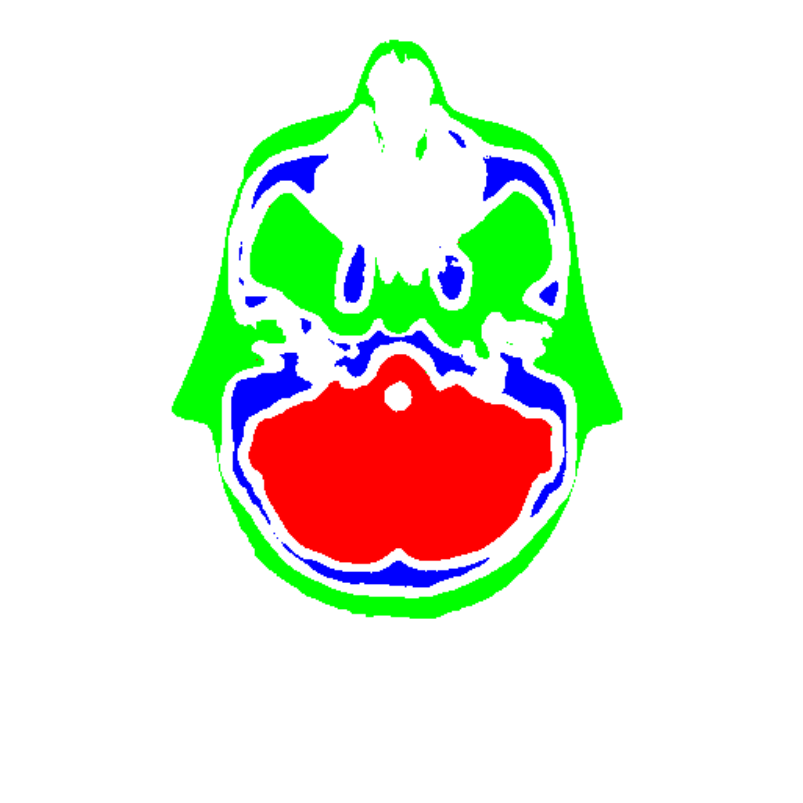}
		\caption{Tissue ROIs}
	\end{subfigure}
	\caption{Illustrations of reconstructed volumes of STEEV phantom with display window using slice 77 [0.7,1.4].}
	\label{fig4:steev_recon}
\end{figure}

Reconstructions of the head phantom are shown in Figure~\ref{fig4:steev_recon}. Observations are that from the `no estimate' in Figure~\ref{fig4:steev_recon}a exhibits similar shading artifacts in the soft tissue regions as the simulated head case in Figure~\ref{fig4:head_sim_recon}b. Between the two scatter estimation approaches, both appear to similarly correct for inhomogeneities from scatter, though the Pre-fASKS does show higher intensities than PolySKS. Although we are unsure of the exact reason why fASKS leads to a consistent overestimation of scatter in our real CBCT data experiments, we should stress that the method is designed only to enhance uniformity, subject to a subsequent calibration of the image, unlike our approach of direct quantification.

To quantitatively assess these reconstructions, we calculated the mean and standard deviation of homogeneous tissue types: brain, soft tissue and soft bone. We isolate regions from these materials with a simple threshold based segmentation of the Pre-fASKS reconstruction, and have shown the colored regions in green, red and blue respectively in Figure~\ref{fig4:steev_recon}d. These are then reported in Table~\ref{tab4:steev_recon}.

\begin{table}[!htb]
	\caption{Quantitative results on soft tissues slice 77 of STEEV head phantom.}
	\label{tab4:steev_recon}
	\centering
	$
	\begin{array}{c|c|c|c|c|c|c|c}
	\text{Material} & \multicolumn{2}{c|}{\text{Soft Tissue}}&\multicolumn{2}{c|}{\text{Brain}}&\multicolumn{2}{c|}{\text{Soft Bone}}& \multicolumn{1}{c}{\text{Total}}\\
	\hline
	\text{ROI color} & \multicolumn{2}{c|}{\text{green}}&\multicolumn{2}{c|}{\text{red}}&\multicolumn{2}{c|}{\text{blue}}& \multicolumn{1}{c}{\text{all}}\\
	\rho_e & \multicolumn{2}{c|}{\text{1.028}}&\multicolumn{2}{c|}{\text{1.039}}&\multicolumn{2}{c|}{\text{1.157}}& \multicolumn{1}{c}{-}\\
	\hline
	& \text{\% mean} & \text{RMSE} & \text{\% mean} & \text{RMSE} & \text{\% mean} & \text{RMSE}& \text{RMSE}\\
	\hline
	\text{No estimate} & -2.26 & 0.0430 & -1.41 & 0.0158 & -3.14 & 0.0409 & 0.0345 \\
	\text{Pre-fASKS} & 1.44 & \mathbf{0.0201} & 1.57 & 0.0173 & 3.92 & 0.0478 & 0.0253 \\
	\text{PolySKS} & \mathbf{0.0544} & 0.0209 & \mathbf{0.439} & \mathbf{0.00709} & \mathbf{0.687} & \mathbf{0.0169} & \mathbf{0.0162} \\
	\hline
	\end{array}
	$
\end{table}

From the quantitative results in Table~\ref{tab4:steev_recon} it can be seen that similarly to the insert phantom, the PolySKS scatter integration leads to the most quantitatively accurate reconstruction. Also similar to the insert case, is the trend with the no estimate leading to an underestimation of electron density, and Pre-fASKS overcompensating to an over estimation. With the exception of the RMSE on the soft tissue, the PolySKS is the best performing method in every metric. In this case however, the Pre-fASKS is the best performing approach due to a lower variation throughout this tissue. It could be that one of the additional correction factors such as the asymmetric modulation, detector glare or couch scatter, not included in our tested PolySKS implementation, may be worthwhile including in our approach also. We also expect the parameters in Table~\ref{tab:parameters} may be improved through fitting to real measurements of scatter. Even with this discrepancy though, PolySKS shows a significant advantage over the whole image, with a 36\% drop in total RMSE over Pre-fASKS.

\section{Discussion} \label{sec4:discussion}
There a several important aspects of our method, about which we would like to discuss: its extension with other perturbations, the computational impact of the new scatter model, and its use in other models besides Polyquant.

\subsection{Extension of the PolySKS Model}
Due to the non-exactness of convolutional scatter estimations, there are many heuristics that may enhance their practical performance under certain conditions. In our presentation of PolySKS, we have tried to present a simple model, whilst employing a couple of empirical perturbations that were found to give significant performance gains for minimal computational cost. Naturally, one could employ several other extensions also, and we will discuss some of these options.

Firstly, we have not included the asymmetric modulation of fASKS in our model --- active when $\gamma>0$ in (\ref{equ:fasks}). This could certainly be applied both on the total scatter factors, or separately on each energy signal. We have performed preliminary experiments in both of these cases, but did not find any significant gain in performance for our numerical data, whilst increasing the computational cost by at least $2\times$. Instead, we found just the edge factor derived in Section~\ref{sec:edge_effects} gave good compensation of inhomogeneities with our data. It could be that objects with more complicated variations in thickness such as the torso, with arms and lung regions may benefit from its inclusion.

Another factor that may be worth considering is the effect of objects outside the field of view, such as the patient couch or scatter from other hardware. In \citet{Sun2011}, the authors do report on this having significant effects on scatter estimation, and its inclusion may widen the gap between PolySKS and the commercial fASKS, where this is employed, in our experimental section. In our numerical tests however, we do not have any additional hardware, allowing us to fairly characterize the relative effectiveness of the methods under test.

We have focussed on convolutional scatter estimation due to their quick run time with FFT implementations. However, a new method using the linear Boltzmann transport equation promises both speed and accuracy \citep{Maslowski2018,Wang2018}. Due to the wealth of information available during the iterates of Polyquant, we believe this method may be readily integrated into the algorithm as with Polyquant--PolySKS, but the gain this may offer remains to be demonstrated. In any case, it has been shown that a similar model may be used to perturb convolutional models and increase their accuracy in \citet{Sun2014}, which may also offer benefit to our method.

Another potential avenue of research is attempting to apply convolutional neural network learning to the polyenergetic scatter factor developed here. A related approach was recently published by \citet{Maier2018} during the development of this work, but this was a preprocessing correction for monoenergetic assumptions. It is possible that such a network could adapt to complex scattering structures, instead of opting for simple perturbations from a slab model, such as the edge factor in Section~\ref{sec:edge_effects}.

\subsection{Comparison with Hardware Scatter Estimation}
An advantage of convolutional scatter estimation is that it requires no additional hardware, and allows use of the full detector for measurements. Beam block solutions to scatter measurement such as collimator shadow or BSA may also offer accurate estimation, and will typically allow much faster processing times as demonstrated in Table~\ref{tab4:sim_results}. 

These two approaches may also be combined. Similarly to the perturbation of fASKS by the Boltzmann transport equation in \citet{Sun2014}, a convolutional estimate may be adjusted to match measurements from beam blocker shadows.

\subsection{Computational Feasibility} \label{comp:feas}
A critical issue with scatter estimation methods is their computational complexity, especially when they are performed at every iteration of reconstruction, such as with Polyquant--PolySKS. Since the motivation behind maintaining a spatially invariant convolutional approach is fast implementation, it is certainly worth directly addressing this point.

As we highlighted in Section~\ref{sec4:poly_scat_model}, calculating the convolution itself bears a cost of $N_\mathrm{proj}$ IFFT operations and $N_\mathrm{proj}(N_\xi+1)$ FFT operations per data pass. Comparatively, the fASKS method has a convolution cost of $2N_\mathrm{proj}N_\mathrm{iter}$ IFFTs and $2N_\mathrm{proj}N_\mathrm{thick}N_\mathrm{iter}$ FFTs, where $N_\mathrm{thick}$ are the number of thickness groups and $N_\mathrm{iter}$ are the number of iterations, which are set to 3 and 4 respectively in \cite{Sun2010}. If one set $N_\mathrm{thick} = 3$ and $N_\xi = 21$ as we have in our experiments, then we note that one could only perform 1.4 epochs for the same cost of fASKS. Therefore, for the 50 data passes we use in our real data experiments, PolySKS carries a notably higher cost. However, we believe that sub-sampling the energy bins is unlikely to have a vast impact of its accuracy, whilst providing big speed ups.

Although when compared to fASKS our estimation approach is likely to be more expensive, we highlight the fact that relative to the cost of performing iterative reconstruction, it is still reasonable. In both the simulated and real data tests, the practical additional cost was 60\%, with no attempt to sub-sample any of the scatter calculation. Our implementation made in \textsc{Matlab} is also likely to be sped up significantly by rewriting in a compiled language such as C++, for which \cite{Sun2010} saw a $4.7\times$ speed up for a single thread with their fASKS --- an implementation on GPGPU hardware is likely to be many times faster still. Additionally, over methods such as Monte Carlo or single scatter models, PolySKS is still likely to be many times faster.

\subsection{Application of PolySKS in Other Methods}
Although we have developed this scatter model specifically for the Polyquant framework, it is certainly available for alternative methods also. Firstly, due to the reported advantage of Int-fASKS over Pre-fASKS, one may use this scatter integration concept with any iterative method, and expect to realize a significant gain by also fusing the scatter estimation with reconstruction.

With the PolySKS scatter model, we believe it may be incorporated into most polyenergetic likelihood models, though with concessions. For methods that quantify a monoenergetic attenuation or mass density rather than electron density such as IMPACT \citep{DeMan2001a} or Poly-SIR respectively \citep{Elbakri2002}, the information $\boldsymbol{\Phi\rho}_e$ is not available at each iteration. To overcome this, one could either perform a non linear conversion from the projection information $\boldsymbol{\Phi\mu}(\xi)$, or indeed use directly $\boldsymbol{\Phi\mu}(\xi)$ in place of $\boldsymbol{\Phi\rho}_e$. This would in fact be directly proportional for water, with the constant of proportionality being its mass attenuation coefficient, but is likely to lose accuracy over diverse material classes.

As a preprocessing method before an analytic reconstruction such as FDK \citep{Feldkamp1984} however, the polyenergetic convolutional model can not be directly applied, due to not having an energy decomposition of the attenuation available. One possible way to exploit it though may be to use an approximate water--bone beam hardening model such as \cite{Joseph1978} to approximate the spectral properties. Additionally, from dual energy (or spectral) CT measurements, such a decomposition may be inferred and the model may be applicable. In both cases, one would likely have to apply the model iteratively as a deconvolution of the contaminated primary measurements as in Pre-fASKS \citep{Sun2010} and it is not clear this would offer any gain.

\section{Conclusions}
We have introduced a polyenergetic convolutional scatter model, and integrated it into a quantitative iterative reconstruction method, denoted Polyquant--PolySKS. Additionally, have developed a fast algorithm for its implementation. The advantages of this are twofold: we have demonstrated through using the same convolutional model, fASKS, that one achieves better reconstruction performance when integrated rather than used in preprocessing; and we have demonstrated that one can exploit polyenergetic information available during Polyquant to gain a higher accuracy still. Due to its spatial invariance, the scatter calculation can be done rapidly through FFT operations, and contributes only a fraction of additional computational time over an iterative reconstruction. From our experiments with both numerical and real data, we have demonstrated superior electron density accuracy with this method, which promises to enhance quantitative medical procedures such as radiation therapy dose calculation.

\section*{Acknowledgments}
The authors would like to sincerely thank Dr Adam Wang from Varian Medical Systems, for preprocessing our CBCT data, providing invaluable information and advice about the imaging system, and making the CIRS insert phantom data available to us. We would also like to thank Andiappa Sankar for his assistance in performing the CBCT acquisition. This work was supported by the Maxwell Advanced Technology Fund, EPSRC DTP studentship funds and ERC project: C-SENSE (ERC-ADG-2015-694888). MD is also supported by a Royal Society Wolfson Research Merit Award.

\bibliographystyle{apalike}
\bibliography{thesis}
\end{document}